\documentclass[submission,copyright,creativecommons]{eptcs}
\providecommand{\event}{ThEdu'24 post-proceedings} %

\usepackage{iftex}

\ifpdf
  \usepackage{underscore}         %
  \usepackage[T1]{fontenc}        %
\else
  \usepackage{breakurl}           %
\fi

\usepackage{academicons}   %
\usepackage[utf8x]{inputenc}
\usepackage[english]{babel} 

\usepackage{amsmath}
\usepackage{amssymb}
\usepackage{pdflscape}
\usepackage{framed}

\usepackage{orcidlink}
\usepackage{graphicx}
\usepackage{float}
\usepackage{subcaption}

\usepackage{xcolor}

\usepackage{hyperref}
 \hypersetup{colorlinks = true,
              citecolor = teal,
              urlcolor = purple}
\usepackage{url}
\usepackage{enumitem}    %
\setlist[itemize]{nosep}  %
\setlist[enumerate]{nosep}  %

\usepackage{textcomp} %
\usepackage{listings}
\usepackage{color}
\definecolor{keywordcolor}{rgb}{0.7, 0.1, 0.1}   %
\definecolor{tacticcolor}{rgb}{0.0, 0.1, 0.6}    %
\definecolor{commentcolor}{rgb}{0.4, 0.4, 0.4}   %
\definecolor{symbolcolor}{rgb}{0.0, 0.1, 0.6}    %
\definecolor{sortcolor}{rgb}{0.1, 0.5, 0.1}      %
\definecolor{attributecolor}{rgb}{0.7, 0.1, 0.1} %
\definecolor{dkgreen}{rgb}{0.1, 0.5, 0.1}      %
\definecolor{dkviolet}{rgb}{0.1, 0.5, 0.1}      %
\definecolor{dkblue}{rgb}{0.0, 0.1, 0.6}    %
\definecolor{dkred}{rgb}{0.0, 0.1, 0.6}    %
\definecolor{ltblue}{rgb}{0.0, 0.1, 0.6}    %
\definecolor{isarblue}{HTML}{006699}
\definecolor{isargreen}{HTML}{009966}

\lstdefinestyle{coq}{
   belowcaptionskip=1\baselineskip,
  breaklines=true,
  xleftmargin=\parindent,
  language=Coq,
  showstringspaces=false,
  basicstyle=\footnotesize\ttfamily,
  identifierstyle=\color{blue},
  stringstyle=\color{orange},
  frame=single,
}

\lstdefinelanguage{Coq}{ 
    mathescape=true,
    texcl=false, 
    morekeywords=[1]{Section, Module, End, Require, Import, Export,
        Variable, Variables, Parameter, Parameters, Axiom, Hypothesis,
        Hypotheses, Notation, Local, Tactic, Reserved, Scope, Open, Close,
        Bind, Delimit, Definition, Let, Ltac, Fixpoint, CoFixpoint, Add,
        Morphism, Relation, Implicit, Arguments, Unset, Contextual,
        Strict, Prenex, Implicits, Inductive, CoInductive, Record,
        Structure, Canonical, Coercion, Context, Class, Global, Instance,
        Program, Infix, Theorem, Lemma, Corollary, Proposition, Fact,
        Remark, Example, Proof, Goal, Save, Qed, Defined, Hint, Resolve,
        Rewrite, View, Search, Show, Print, Printing, All, Eval, Check,
        Projections, inside, outside, Def},
    morekeywords=[2]{forall, exists, exists2, fun, fix, cofix, struct,
        match, with, end, as, in, return, let, if, is, then, else, for, of,
        nosimpl, when},
    morekeywords=[3]{Type, Prop, Set, true, false, option},
    morekeywords=[4]{pose, set, move, case, elim, apply, clear, hnf,
        intro, intros, constructor, generalize, rename, pattern, after, destruct,
        induction, using, refine, inversion, injection, rewrite, congr,
        unlock, compute, ring, field, fourier, replace, fold, unfold,
        change, cutrewrite, simpl, have, suff, wlog, suffices, without,
        loss, nat_norm, assert, cut, trivial, revert, bool_congr, nat_congr,
        symmetry, transitivity, auto, split, left, right, autorewrite},
    morekeywords=[5]{by, done, exact, reflexivity, tauto, romega, omega,
        assumption, solve, contradiction, discriminate},
    morekeywords=[6]{do, last, first, try, idtac, repeat},
    morecomment=[s]{(*}{*)},
    showstringspaces=false,
    morestring=[b]",
    morestring=[d]’,
    tabsize=3,
    extendedchars=false,
    sensitive=true,
    breaklines=false,
    basicstyle=\small,
    captionpos=b,
    columns=[l]flexible,
    identifierstyle={\ttfamily\color{black}},
    keywordstyle=[1]{\ttfamily\color{keywordcolor}},
    keywordstyle=[2]{\ttfamily\color{black}},
    keywordstyle=[3]{\ttfamily\color{sortcolor}},
    keywordstyle=[4]{\ttfamily\color{tacticcolor}},
    keywordstyle=[5]{\ttfamily\color{dkred}},
    stringstyle=\ttfamily,
    commentstyle={\ttfamily\color{commentcolor}},
    literate=
    {\\forall}{{{$\forall\;$}}}1
    {\\exists}{{$\exists\;$}}1
    {<-}{{$\leftarrow\;$}}1
    {=>}{{$\Rightarrow\;$}}1
    {==}{{\code{==}\;}}1
    {==>}{{\code{==>}\;}}1
    {->}{{$\rightarrow\;$}}1
    {<->}{{$\leftrightarrow\;$}}1
    {<==}{{$\leq\;$}}1
    {\#}{{$^\star$}}1 
    {\\o}{{$\circ\;$}}1 
    {\@}{{$\cdot$}}1 
    {\/\\}{{$\wedge\;$}}1
    {\\\/}{{$\vee\;$}}1
    {++}{{\code{++}}}1
    {~}{{$\sim$}}1
    {\@\@}{{$@$}}1
    {\\mapsto}{{$\mapsto\;$}}1
    {\\hline}{{\rule{\linewidth}{0.5pt}}}1
{à}{{\`a}}1
{è}{{\`e}}1
{ù}{{\`u}}1
{á}{{\'a}}1
{é}{{\'e}}1
{ú}{{\'u}}1
{Á}{{\'A}}1
{É}{{\'E}}1
{Ú}{{\'U}}1
{ô}{{\^o}}1
{ê}{{\^e}}1
{â}{{\^a}}1
{ç}{{\c c}}1         
{ℓ}{{\Ell}}1
{α}{{\ensuremath{\mathrm{\alpha}}}}1
{β}{{\ensuremath{\mathrm{\beta}}}}1
{γ}{{\ensuremath{\mathrm{\gamma}}}}1
{δ}{{\ensuremath{\mathrm{\delta}}}}1
{ε}{{\ensuremath{\mathrm{\varepsilon}}}}1
{ζ}{{\ensuremath{\mathrm{\zeta}}}}1
{η}{{\ensuremath{\mathrm{\eta}}}}1
{θ}{{\ensuremath{\mathrm{\theta}}}}1
{ι}{{\ensuremath{\mathrm{\iota}}}}1
{κ}{{\ensuremath{\mathrm{\kappa}}}}1
{μ}{{\ensuremath{\mathrm{\mu}}}}1
{ν}{{\ensuremath{\mathrm{\nu}}}}1
{ξ}{{\ensuremath{\mathrm{\xi}}}}1
{π}{{\ensuremath{\mathrm{\mathnormal{\pi}}}}}1
{ρ}{{\ensuremath{\mathrm{\rho}}}}1
{σ}{{\ensuremath{\mathrm{\sigma}}}}1
{τ}{{\ensuremath{\mathrm{\tau}}}}1
{φ}{{\ensuremath{\mathrm{\varphi}}}}1
{χ}{{\ensuremath{\mathrm{\chi}}}}1
{ψ}{{\ensuremath{\mathrm{\psi}}}}1
{ω}{{\ensuremath{\mathrm{\omega}}}}1
{Γ}{{\ensuremath{\mathrm{\Gamma}}}}1
{Δ}{{\ensuremath{\mathrm{\Delta}}}}1
{Θ}{{\ensuremath{\mathrm{\Theta}}}}1
{Λ}{{\ensuremath{\mathrm{\Lambda}}}}1
{Σ}{{\ensuremath{\mathrm{\Sigma}}}}1
{Φ}{{\ensuremath{\mathrm{\Phi}}}}1
{Ξ}{{\ensuremath{\mathrm{\Xi}}}}1
{Ψ}{{\ensuremath{\mathrm{\Psi}}}}1
{Ω}{{\ensuremath{\mathrm{\Omega}}}}1
{ℵ}{{\ensuremath{\aleph}}}1
{≤}{{\ensuremath{\leq}}}1
{≥}{{\ensuremath{\geq}}}1
{≠}{{\ensuremath{\neq}}}1
{≈}{{\ensuremath{\approx}}}1
{≡}{{\ensuremath{\equiv}}}1
{≃}{{\ensuremath{\simeq}}}1
{∂}{{\ensuremath{\partial}}}1
{∆}{{\ensuremath{\triangle}}}1 
{∫}{{\ensuremath{\int}}}1
{∑}{{\ensuremath{\mathrm{\Sigma}}}}1
{Π}{{\ensuremath{\mathrm{\Pi}}}}1
{⊥}{{\ensuremath{\perp}}}1
{∞}{{\ensuremath{\infty}}}1
{∂}{{\ensuremath{\partial}}}1
{∓}{{\ensuremath{\mp}}}1
{±}{{\ensuremath{\pm}}}1
{×}{{\ensuremath{\times}}}1
{⊕}{{\ensuremath{\oplus}}}1
{⊗}{{\ensuremath{\otimes}}}1
{⊞}{{\ensuremath{\boxplus}}}1
{∇}{{\ensuremath{\nabla}}}1
{√}{{\ensuremath{\sqrt}}}1
{⬝}{{\ensuremath{\cdot}}}1
{•}{{\ensuremath{\cdot}}}1
{∘}{{\ensuremath{\circ}}}1
{⁻}{{\ensuremath{^{-}}}}1
{▸}{{\ensuremath{\blacktriangleright}}}1
{∧}{{\ensuremath{\wedge}}}1
{∨}{{\ensuremath{\vee}}}1
{¬}{{\ensuremath{\neg}}}1
{⊢}{{\ensuremath{\vdash}}}1
{⟨}{{\ensuremath{\langle}}}1
{⟩}{{\ensuremath{\rangle}}}1
{↦}{{\ensuremath{\mapsto}}}1
{←}{{\ensuremath{\leftarrow}}}1
{<-}{{\ensuremath{\leftarrow}}}1
{→}{{\ensuremath{\rightarrow}}}1
{⟶}{{\ensuremath{\longrightarrow}}}1    
{↔}{{\ensuremath{\leftrightarrow}}}1
{⟷}{{\ensuremath{\longleftrightarrow}}}1    
{⇒}{{\ensuremath{\Rightarrow}}}1
{⟹}{{\ensuremath{\Longrightarrow}}}1
{⇐}{{\ensuremath{\Leftarrow}}}1
{⟸}{{\ensuremath{\Longleftarrow}}}1
{ℕ}{{\ensuremath{\mathbb{N}}}}1
{ℤ}{{\ensuremath{\mathbb{Z}}}}1
{ℝ}{{\ensuremath{\mathbb{R}}}}1
{ℚ}{{\ensuremath{\mathbb{Q}}}}1
{ℂ}{{\ensuremath{\mathbb{C}}}}1
{⌞}{{\ensuremath{\llcorner}}}1
{⌟}{{\ensuremath{\lrcorner}}}1
{⦃}{{\ensuremath{\{\!|}}}1
{⦄}{{\ensuremath{|\!\}}}}1
{‖}{{\ensuremath{\|}}}1
{₁}{{\ensuremath{_1}}}1
{₂}{{\ensuremath{_2}}}1
{₃}{{\ensuremath{_3}}}1
{₄}{{\ensuremath{_4}}}1
{₅}{{\ensuremath{_5}}}1
{₆}{{\ensuremath{_6}}}1
{₇}{{\ensuremath{_7}}}1
{₈}{{\ensuremath{_8}}}1
{₉}{{\ensuremath{_9}}}1
{₀}{{\ensuremath{_0}}}1
{ᵢ}{{\ensuremath{_i}}}1
{ⱼ}{{\ensuremath{_j}}}1
{ₐ}{{\ensuremath{_a}}}1
{¹}{{\ensuremath{^1}}}1
{ₙ}{{\ensuremath{_n}}}1
{ₘ}{{\ensuremath{_m}}}1
{ₚ}{{\ensuremath{_p}}}1
{↑}{{\ensuremath{\uparrow}}}1
{↓}{{\ensuremath{\downarrow}}}1
{...}{{\ensuremath{\ldots}}}1
{·}{{\ensuremath{\cdot}}}1
}[keywords,comments,strings]

\lstdefinestyle{flatsmall}{
   belowcaptionskip=1\baselineskip,
  breaklines=true,
  xleftmargin=\parindent,
  language={},
  showstringspaces=false,
  basicstyle=\tiny\ttfamily,
  keywordstyle=\color{black},
  commentstyle=\color{black},
  identifierstyle=\color{black},
  stringstyle=\color{black},
}

\lstdefinestyle{lean}{
  belowcaptionskip=1\baselineskip,
  breaklines=true,
  xleftmargin=\parindent,
  language=lean,
  showstringspaces=false,
  basicstyle=\footnotesize\ttfamily,
  identifierstyle=\color{blue},
  stringstyle=\color{orange},
  frame=single,
}

\lstdefinelanguage{lean}{
mathescape=false,
texcl=false,
morekeywords=[1]{ import, prelude, protected, private, noncomputable, definition, meta, renaming, hiding, parameter, parameters, begin, constant, constants,
 lemma, variable, variables, theory,
 print, theorem, example,
 open, as, export, override, axiom, axioms, inductive, with,
 structure, record, universe, universes,
 alias, help, precedence, reserve, declare_trace, add_key_equivalence,
 match, infix, infixl, infixr, notation, postfix, prefix, instance,
 eval, reduce, check, end, this,
 using, using_well_founded, namespace, section,
 attribute, local, set_option, extends, include, omit, class,
 raw, replacing,
 calc, have, show, suffices, by, in, at, let, forall, Pi, fun,
 exists, if, dif, then, else, assume, obtain, from, register_simp_ext, unless, break, continue,
 mutual, do, def, run_cmd, const,
 partial, mut, where, macro, syntax, deriving,
 return, try, catch, for, macro_rules, declare_syntax_cat, abbrev},
morekeywords=[2]{Sort, Type, Prop},
morekeywords=[3]{
 assumption, apply, intro, intros, allGoals, generalize, clear, revert, done, exact,
 refine, repeat, cases, rewrite, rw, simp, simp_all, contradiction,
 constructor, injection, induction
 },
literate=
{à}{{\`a}}1
{è}{{\`e}}1
{ù}{{\`u}}1
{á}{{\'a}}1
{é}{{\'e}}1
{ú}{{\'u}}1
{Á}{{\'A}}1
{É}{{\'E}}1
{Ú}{{\'U}}1
{ô}{{\^o}}1
{ê}{{\^e}}1
{â}{{\^a}}1
{ç}{{\c c}}1         
{ℓ}{{\Ell}}1
{α}{{\ensuremath{\mathrm{\alpha}}}}1
{β}{{\ensuremath{\mathrm{\beta}}}}1
{γ}{{\ensuremath{\mathrm{\gamma}}}}1
{δ}{{\ensuremath{\mathrm{\delta}}}}1
{ε}{{\ensuremath{\mathrm{\varepsilon}}}}1
{ζ}{{\ensuremath{\mathrm{\zeta}}}}1
{η}{{\ensuremath{\mathrm{\eta}}}}1
{θ}{{\ensuremath{\mathrm{\theta}}}}1
{ι}{{\ensuremath{\mathrm{\iota}}}}1
{κ}{{\ensuremath{\mathrm{\kappa}}}}1
{μ}{{\ensuremath{\mathrm{\mu}}}}1
{ν}{{\ensuremath{\mathrm{\nu}}}}1
{ξ}{{\ensuremath{\mathrm{\xi}}}}1
{π}{{\ensuremath{\mathrm{\mathnormal{\pi}}}}}1
{ρ}{{\ensuremath{\mathrm{\rho}}}}1
{σ}{{\ensuremath{\mathrm{\sigma}}}}1
{τ}{{\ensuremath{\mathrm{\tau}}}}1
{φ}{{\ensuremath{\mathrm{\varphi}}}}1
{χ}{{\ensuremath{\mathrm{\chi}}}}1
{ψ}{{\ensuremath{\mathrm{\psi}}}}1
{ω}{{\ensuremath{\mathrm{\omega}}}}1
{Γ}{{\ensuremath{\mathrm{\Gamma}}}}1
{Δ}{{\ensuremath{\mathrm{\Delta}}}}1
{Θ}{{\ensuremath{\mathrm{\Theta}}}}1
{Λ}{{\ensuremath{\mathrm{\Lambda}}}}1
{Σ}{{\ensuremath{\mathrm{\Sigma}}}}1
{Φ}{{\ensuremath{\mathrm{\Phi}}}}1
{Ξ}{{\ensuremath{\mathrm{\Xi}}}}1
{Ψ}{{\ensuremath{\mathrm{\Psi}}}}1
{Ω}{{\ensuremath{\mathrm{\Omega}}}}1
{ℵ}{{\ensuremath{\aleph}}}1
{≤}{{\ensuremath{\leq}}}1
{≥}{{\ensuremath{\geq}}}1
{≠}{{\ensuremath{\neq}}}1
{≈}{{\ensuremath{\approx}}}1
{≡}{{\ensuremath{\equiv}}}1
{≃}{{\ensuremath{\simeq}}}1
{∂}{{\ensuremath{\partial}}}1
{∆}{{\ensuremath{\triangle}}}1 
{∫}{{\ensuremath{\int}}}1
{∑}{{\ensuremath{\mathrm{\Sigma}}}}1
{Π}{{\ensuremath{\mathrm{\Pi}}}}1
{⊥}{{\ensuremath{\perp}}}1
{∞}{{\ensuremath{\infty}}}1
{∂}{{\ensuremath{\partial}}}1
{∓}{{\ensuremath{\mp}}}1
{±}{{\ensuremath{\pm}}}1
{×}{{\ensuremath{\times}}}1
{⊕}{{\ensuremath{\oplus}}}1
{⊗}{{\ensuremath{\otimes}}}1
{⊞}{{\ensuremath{\boxplus}}}1
{∇}{{\ensuremath{\nabla}}}1
{√}{{\ensuremath{\sqrt}}}1
{⬝}{{\ensuremath{\cdot}}}1
{•}{{\ensuremath{\cdot}}}1
{∘}{{\ensuremath{\circ}}}1
{⁻}{{\ensuremath{^{-}}}}1
{▸}{{\ensuremath{\blacktriangleright}}}1
{∧}{{\ensuremath{\wedge}}}1
{∨}{{\ensuremath{\vee}}}1
{¬}{{\ensuremath{\neg}}}1
{⊢}{{\ensuremath{\vdash}}}1
{⟨}{{\ensuremath{\langle}}}1
{⟩}{{\ensuremath{\rangle}}}1
{↦}{{\ensuremath{\mapsto}}}1
{←}{{\ensuremath{\leftarrow}}}1
{<-}{{\ensuremath{\leftarrow}}}1
{→}{{\ensuremath{\rightarrow}}}1
{⟶}{{\ensuremath{\longrightarrow}}}1    
{↔}{{\ensuremath{\leftrightarrow}}}1
{⟷}{{\ensuremath{\longleftrightarrow}}}1    
{⇒}{{\ensuremath{\Rightarrow}}}1
{⟹}{{\ensuremath{\Longrightarrow}}}1
{⇐}{{\ensuremath{\Leftarrow}}}1
{⟸}{{\ensuremath{\Longleftarrow}}}1
{∩}{{\ensuremath{\cap}}}1
{∪}{{\ensuremath{\cup}}}1
{⊂}{{\ensuremath{\subseteq}}}1
{⊆}{{\ensuremath{\subseteq}}}1
{⊄}{{\ensuremath{\nsubseteq}}}1
{⊈}{{\ensuremath{\nsubseteq}}}1
{⊃}{{\ensuremath{\supseteq}}}1
{⊇}{{\ensuremath{\supseteq}}}1
{⊅}{{\ensuremath{\nsupseteq}}}1
{⊉}{{\ensuremath{\nsupseteq}}}1
{∈}{{\ensuremath{\in}}}1
{∉}{{\ensuremath{\notin}}}1
{∋}{{\ensuremath{\ni}}}1
{∌}{{\ensuremath{\notni}}}1
{∅}{{\ensuremath{\emptyset}}}1
{∖}{{\ensuremath{\setminus}}}1
{†}{{\ensuremath{\dag}}}1
{ℕ}{{\ensuremath{\mathbb{N}}}}1
{ℤ}{{\ensuremath{\mathbb{Z}}}}1
{ℝ}{{\ensuremath{\mathbb{R}}}}1
{ℚ}{{\ensuremath{\mathbb{Q}}}}1
{ℂ}{{\ensuremath{\mathbb{C}}}}1
{⌞}{{\ensuremath{\llcorner}}}1
{⌟}{{\ensuremath{\lrcorner}}}1
{⦃}{{\ensuremath{\{\!|}}}1
{⦄}{{\ensuremath{|\!\}}}}1
{‖}{{\ensuremath{\|}}}1
{₁}{{\ensuremath{_1}}}1
{₂}{{\ensuremath{_2}}}1
{₃}{{\ensuremath{_3}}}1
{₄}{{\ensuremath{_4}}}1
{₅}{{\ensuremath{_5}}}1
{₆}{{\ensuremath{_6}}}1
{₇}{{\ensuremath{_7}}}1
{₈}{{\ensuremath{_8}}}1
{₉}{{\ensuremath{_9}}}1
{₀}{{\ensuremath{_0}}}1
{ᵢ}{{\ensuremath{_i}}}1
{ⱼ}{{\ensuremath{_j}}}1
{ₐ}{{\ensuremath{_a}}}1
{¹}{{\ensuremath{^1}}}1
{ₙ}{{\ensuremath{_n}}}1
{ₘ}{{\ensuremath{_m}}}1
{ₚ}{{\ensuremath{_p}}}1
{↑}{{\ensuremath{\uparrow}}}1
{↓}{{\ensuremath{\downarrow}}}1
{...}{{\ensuremath{\ldots}}}1
{·}{{\ensuremath{\cdot}}}1
{Σ}{{\color{symbolcolor}\ensuremath{\Sigma}}}1
{Π}{{\color{symbolcolor}\ensuremath{\Pi}}}1
{∀}{{\color{symbolcolor}\ensuremath{\forall}}}1
{∃}{{\color{symbolcolor}\ensuremath{\exists}}}1
{λ}{{\color{symbolcolor}\ensuremath{\mathrm{\lambda}}}}1
{\$}{{\color{symbolcolor}\$}}1
{:=}{{\color{symbolcolor}:=}}1
{=}{{\color{symbolcolor}=}}1
{<|>}{{\color{symbolcolor}<|>}}1
{<\$>}{{\color{symbolcolor}<\$>}}1
{+}{{\color{symbolcolor}+}}1
{*}{{\color{symbolcolor}*}}1,
morecomment=[s][\color{commentcolor}]{/-}{-/},
morecomment=[l][\itshape \color{commentcolor}]{--},
showstringspaces=false,
keepspaces=true,
morestring=[b]",
morestring=[d],
tabsize=3,
extendedchars=false,
sensitive=true,
breaklines=true,
breakatwhitespace=true,
basicstyle=\ttfamily\small,
captionpos=b,
columns=[l]fullflexible,
identifierstyle={\ttfamily\color{black}},
keywordstyle=[1]{\ttfamily\color{keywordcolor}},
keywordstyle=[2]{\ttfamily\color{sortcolor}},
keywordstyle=[3]{\ttfamily\color{tacticcolor}},
keywordstyle=[4]{\ttfamily\color{attributecolor}},
stringstyle=\ttfamily,
commentstyle={\ttfamily\color{commentcolor}}
}

\lstdefinelanguage{leanverbose}{
mathescape=false,
texcl=false,
morekeywords=[1]{ Exercise, Given, Assume, Conclusion, Proof, QED, import, prelude, protected, private, noncomputable, definition, meta, renaming, hiding, parameter, parameters, begin, constant, constants,
 lemma, variable, variables, theory,
 print, theorem, example,
 open, as, export, override, axiom, axioms, inductive, with,
 structure, record, universe, universes,
 alias, help, precedence, reserve, declare_trace, add_key_equivalence,
 match, infix, infixl, infixr, notation, postfix, prefix, instance,
 eval, reduce, check, end, this,
 using_well_founded, namespace, section,
 attribute, local, set_option, extends, include, omit, class,
 raw, replacing,
 calc, have, show, in, at, let, forall, Pi, fun,
 exists, if, dif, then, else, assume, obtain, from, register_simp_ext, unless, break, continue,
 mutual, do, def, run_cmd, const,
 partial, mut, where, macro, syntax, deriving,
 return, try, catch, for, macro_rules, declare_syntax_cat, abbrev},
morekeywords=[2]{Sort, Type, Prop},
morekeywords=[3]{
 Let, prove, Fix, By, applied, to, that, using, suffices, works, We, we, such, and, 's, by,  get, conclude, it,
 assumption, apply, intro, intros, allGoals, generalize, clear, revert, done, exact,
 refine, repeat, cases, rewrite, rw, simp, simp_all, contradiction,
 constructor, injection, induction
 },
literate=
{à}{{\`a}}1
{è}{{\`e}}1
{ù}{{\`u}}1
{á}{{\'a}}1
{é}{{\'e}}1
{ú}{{\'u}}1
{Á}{{\'A}}1
{É}{{\'E}}1
{Ú}{{\'U}}1
{ô}{{\^o}}1
{ê}{{\^e}}1
{â}{{\^a}}1
{ç}{{\c c}}1         
{ℓ}{{\Ell}}1
{α}{{\ensuremath{\mathrm{\alpha}}}}1
{β}{{\ensuremath{\mathrm{\beta}}}}1
{γ}{{\ensuremath{\mathrm{\gamma}}}}1
{δ}{{\ensuremath{\mathrm{\delta}}}}1
{ε}{{\ensuremath{\mathrm{\varepsilon}}}}1
{ζ}{{\ensuremath{\mathrm{\zeta}}}}1
{η}{{\ensuremath{\mathrm{\eta}}}}1
{θ}{{\ensuremath{\mathrm{\theta}}}}1
{ι}{{\ensuremath{\mathrm{\iota}}}}1
{κ}{{\ensuremath{\mathrm{\kappa}}}}1
{μ}{{\ensuremath{\mathrm{\mu}}}}1
{ν}{{\ensuremath{\mathrm{\nu}}}}1
{ξ}{{\ensuremath{\mathrm{\xi}}}}1
{π}{{\ensuremath{\mathrm{\mathnormal{\pi}}}}}1
{ρ}{{\ensuremath{\mathrm{\rho}}}}1
{σ}{{\ensuremath{\mathrm{\sigma}}}}1
{τ}{{\ensuremath{\mathrm{\tau}}}}1
{φ}{{\ensuremath{\mathrm{\varphi}}}}1
{χ}{{\ensuremath{\mathrm{\chi}}}}1
{ψ}{{\ensuremath{\mathrm{\psi}}}}1
{ω}{{\ensuremath{\mathrm{\omega}}}}1
{Γ}{{\ensuremath{\mathrm{\Gamma}}}}1
{Δ}{{\ensuremath{\mathrm{\Delta}}}}1
{Θ}{{\ensuremath{\mathrm{\Theta}}}}1
{Λ}{{\ensuremath{\mathrm{\Lambda}}}}1
{Σ}{{\ensuremath{\mathrm{\Sigma}}}}1
{Φ}{{\ensuremath{\mathrm{\Phi}}}}1
{Ξ}{{\ensuremath{\mathrm{\Xi}}}}1
{Ψ}{{\ensuremath{\mathrm{\Psi}}}}1
{Ω}{{\ensuremath{\mathrm{\Omega}}}}1
{ℵ}{{\ensuremath{\aleph}}}1
{≤}{{\ensuremath{\leq}}}1
{≥}{{\ensuremath{\geq}}}1
{≠}{{\ensuremath{\neq}}}1
{≈}{{\ensuremath{\approx}}}1
{≡}{{\ensuremath{\equiv}}}1
{≃}{{\ensuremath{\simeq}}}1
{∂}{{\ensuremath{\partial}}}1
{∆}{{\ensuremath{\triangle}}}1 
{∫}{{\ensuremath{\int}}}1
{∑}{{\ensuremath{\mathrm{\Sigma}}}}1
{Π}{{\ensuremath{\mathrm{\Pi}}}}1
{⊥}{{\ensuremath{\perp}}}1
{∞}{{\ensuremath{\infty}}}1
{∂}{{\ensuremath{\partial}}}1
{∓}{{\ensuremath{\mp}}}1
{±}{{\ensuremath{\pm}}}1
{×}{{\ensuremath{\times}}}1
{⊕}{{\ensuremath{\oplus}}}1
{⊗}{{\ensuremath{\otimes}}}1
{⊞}{{\ensuremath{\boxplus}}}1
{∇}{{\ensuremath{\nabla}}}1
{√}{{\ensuremath{\sqrt}}}1
{⬝}{{\ensuremath{\cdot}}}1
{•}{{\ensuremath{\cdot}}}1
{∘}{{\ensuremath{\circ}}}1
{⁻}{{\ensuremath{^{-}}}}1
{▸}{{\ensuremath{\blacktriangleright}}}1
{∧}{{\ensuremath{\wedge}}}1
{∨}{{\ensuremath{\vee}}}1
{¬}{{\ensuremath{\neg}}}1
{⊢}{{\ensuremath{\vdash}}}1
{⟨}{{\ensuremath{\langle}}}1
{⟩}{{\ensuremath{\rangle}}}1
{↦}{{\ensuremath{\mapsto}}}1
{←}{{\ensuremath{\leftarrow}}}1
{<-}{{\ensuremath{\leftarrow}}}1
{→}{{\ensuremath{\rightarrow}}}1
{⟶}{{\ensuremath{\longrightarrow}}}1    
{↔}{{\ensuremath{\leftrightarrow}}}1
{⟷}{{\ensuremath{\longleftrightarrow}}}1    
{⇒}{{\ensuremath{\Rightarrow}}}1
{⟹}{{\ensuremath{\Longrightarrow}}}1
{⇐}{{\ensuremath{\Leftarrow}}}1
{⟸}{{\ensuremath{\Longleftarrow}}}1
{∩}{{\ensuremath{\cap}}}1
{∪}{{\ensuremath{\cup}}}1
{⊂}{{\ensuremath{\subseteq}}}1
{⊆}{{\ensuremath{\subseteq}}}1
{⊄}{{\ensuremath{\nsubseteq}}}1
{⊈}{{\ensuremath{\nsubseteq}}}1
{⊃}{{\ensuremath{\supseteq}}}1
{⊇}{{\ensuremath{\supseteq}}}1
{⊅}{{\ensuremath{\nsupseteq}}}1
{⊉}{{\ensuremath{\nsupseteq}}}1
{∈}{{\ensuremath{\in}}}1
{∉}{{\ensuremath{\notin}}}1
{∋}{{\ensuremath{\ni}}}1
{∌}{{\ensuremath{\notni}}}1
{∅}{{\ensuremath{\emptyset}}}1
{∖}{{\ensuremath{\setminus}}}1
{†}{{\ensuremath{\dag}}}1
{ℕ}{{\ensuremath{\mathbb{N}}}}1
{ℤ}{{\ensuremath{\mathbb{Z}}}}1
{ℝ}{{\ensuremath{\mathbb{R}}}}1
{ℚ}{{\ensuremath{\mathbb{Q}}}}1
{ℂ}{{\ensuremath{\mathbb{C}}}}1
{⌞}{{\ensuremath{\llcorner}}}1
{⌟}{{\ensuremath{\lrcorner}}}1
{⦃}{{\ensuremath{\{\!|}}}1
{⦄}{{\ensuremath{|\!\}}}}1
{‖}{{\ensuremath{\|}}}1
{₁}{{\ensuremath{_1}}}1
{₂}{{\ensuremath{_2}}}1
{₃}{{\ensuremath{_3}}}1
{₄}{{\ensuremath{_4}}}1
{₅}{{\ensuremath{_5}}}1
{₆}{{\ensuremath{_6}}}1
{₇}{{\ensuremath{_7}}}1
{₈}{{\ensuremath{_8}}}1
{₉}{{\ensuremath{_9}}}1
{₀}{{\ensuremath{_0}}}1
{ᵢ}{{\ensuremath{_i}}}1
{ⱼ}{{\ensuremath{_j}}}1
{ₐ}{{\ensuremath{_a}}}1
{¹}{{\ensuremath{^1}}}1
{ₙ}{{\ensuremath{_n}}}1
{ₘ}{{\ensuremath{_m}}}1
{ₚ}{{\ensuremath{_p}}}1
{↑}{{\ensuremath{\uparrow}}}1
{↓}{{\ensuremath{\downarrow}}}1
{...}{{\ensuremath{\ldots}}}1
{·}{{\ensuremath{\cdot}}}1
{Σ}{{\color{symbolcolor}\ensuremath{\Sigma}}}1
{Π}{{\color{symbolcolor}\ensuremath{\Pi}}}1
{∀}{{\color{symbolcolor}\ensuremath{\forall}}}1
{∃}{{\color{symbolcolor}\ensuremath{\exists}}}1
{λ}{{\color{symbolcolor}\ensuremath{\mathrm{\lambda}}}}1
{\$}{{\color{symbolcolor}\$}}1
{:=}{{\color{symbolcolor}:=}}1
{=}{{\color{symbolcolor}=}}1
{<|>}{{\color{symbolcolor}<|>}}1
{<\$>}{{\color{symbolcolor}<\$>}}1
{+}{{\color{symbolcolor}+}}1
{*}{{\color{symbolcolor}*}}1,
morecomment=[s][\color{commentcolor}]{/-}{-/},
morecomment=[l][\itshape \color{commentcolor}]{--},
showstringspaces=false,
keepspaces=true,
morestring=[b]",
morestring=[d],
tabsize=3,
extendedchars=false,
sensitive=true,
breaklines=true,
breakatwhitespace=true,
basicstyle=\ttfamily\small,
captionpos=b,
columns=[l]fullflexible,
identifierstyle={\ttfamily\color{black}},
keywordstyle=[1]{\ttfamily\color{keywordcolor}},
keywordstyle=[2]{\ttfamily\color{sortcolor}},
keywordstyle=[3]{\ttfamily\color{tacticcolor}},
keywordstyle=[4]{\ttfamily\color{attributecolor}},
stringstyle=\ttfamily,
commentstyle={\ttfamily\color{commentcolor}}
}
\lstdefinestyle{isabelle}{
   belowcaptionskip=1\baselineskip,
  breaklines=true,
  xleftmargin=\parindent,
  language=isabelle,
  showstringspaces=false,
  basicstyle=\footnotesize\ttfamily,
  identifierstyle=\color{blue},
  stringstyle=\color{orange},
  frame=single,
}

\lstdefinelanguage{isabelle}{%
    keywords=[1]{type_synonym,datatype,fun,abbreviation,definition,proof,lemma,theorem,corollary,qed},
    keywordstyle=[1]\ttfamily\color{keywordcolor},
    keywords=[2]{where,assumes,assume,thus,have,shows,from,and,show,by},
    keywordstyle=[2]\ttfamily\color{keywordcolor},
    keywords=[3]{if,then,else,case,of,SOME,let,in,O},
    keywordstyle=[3]\color{isarblue},
    keywords=[4]{blast,simp,auto,cutting},
    keywordstyle=[4]\color{tacticcolor},
    morecomment=[s]{(*}{*)},
    basicstyle=\small\ttfamily,
    commentstyle={\ttfamily\color{commentcolor}},
    breaklines=true,
    breakatwhitespace=true,
    literate=
    {à}{{\`a}}1
    {è}{{\`e}}1
    {ù}{{\`u}}1
    {á}{{\'a}}1
    {é}{{\'e}}1
    {ú}{{\'u}}1
    {Á}{{\'A}}1
    {É}{{\'E}}1
    {Ú}{{\'U}}1
    {ô}{{\^o}}1
    {ê}{{\^e}}1
    {â}{{\^a}}1
    {ç}{{\c c}}1         
    {ℓ}{{\Ell}}1
    {α}{{\ensuremath{\mathrm{\alpha}}}}1
    {β}{{\ensuremath{\mathrm{\beta}}}}1
    {γ}{{\ensuremath{\mathrm{\gamma}}}}1
    {δ}{{\ensuremath{\mathrm{\delta}}}}1
    {ε}{{\ensuremath{\mathrm{\varepsilon}}}}1
    {ζ}{{\ensuremath{\mathrm{\zeta}}}}1
    {η}{{\ensuremath{\mathrm{\eta}}}}1
    {θ}{{\ensuremath{\mathrm{\theta}}}}1
    {ι}{{\ensuremath{\mathrm{\iota}}}}1
    {κ}{{\ensuremath{\mathrm{\kappa}}}}1
    {μ}{{\ensuremath{\mathrm{\mu}}}}1
    {ν}{{\ensuremath{\mathrm{\nu}}}}1
    {ξ}{{\ensuremath{\mathrm{\xi}}}}1
    {π}{{\ensuremath{\mathrm{\mathnormal{\pi}}}}}1
    {ρ}{{\ensuremath{\mathrm{\rho}}}}1
    {σ}{{\ensuremath{\mathrm{\sigma}}}}1
    {τ}{{\ensuremath{\mathrm{\tau}}}}1
    {φ}{{\ensuremath{\mathrm{\varphi}}}}1
    {χ}{{\ensuremath{\mathrm{\chi}}}}1
    {ψ}{{\ensuremath{\mathrm{\psi}}}}1
    {ω}{{\ensuremath{\mathrm{\omega}}}}1
    {Γ}{{\ensuremath{\mathrm{\Gamma}}}}1
    {Δ}{{\ensuremath{\mathrm{\Delta}}}}1
    {Θ}{{\ensuremath{\mathrm{\Theta}}}}1
    {Λ}{{\ensuremath{\mathrm{\Lambda}}}}1
    {Σ}{{\ensuremath{\mathrm{\Sigma}}}}1
    {Φ}{{\ensuremath{\mathrm{\Phi}}}}1
    {Ξ}{{\ensuremath{\mathrm{\Xi}}}}1
    {Ψ}{{\ensuremath{\mathrm{\Psi}}}}1
    {Ω}{{\ensuremath{\mathrm{\Omega}}}}1
    {ℵ}{{\ensuremath{\aleph}}}1
    {≤}{{\ensuremath{\leq}}}1
    {≥}{{\ensuremath{\geq}}}1
    {≠}{{\ensuremath{\neq}}}1
    {≈}{{\ensuremath{\approx}}}1
    {≡}{{\ensuremath{\equiv}}}1
    {≃}{{\ensuremath{\simeq}}}1
    {∂}{{\ensuremath{\partial}}}1
    {∆}{{\ensuremath{\triangle}}}1 
    {∫}{{\ensuremath{\int}}}1
    {∑}{{\ensuremath{\mathrm{\Sigma}}}}1
    {Π}{{\ensuremath{\mathrm{\Pi}}}}1
    {⊥}{{\ensuremath{\perp}}}1
    {∞}{{\ensuremath{\infty}}}1
    {∂}{{\ensuremath{\partial}}}1
    {∓}{{\ensuremath{\mp}}}1
    {±}{{\ensuremath{\pm}}}1
    {×}{{\ensuremath{\times}}}1
    {⊕}{{\ensuremath{\oplus}}}1
    {⊗}{{\ensuremath{\otimes}}}1
    {⊞}{{\ensuremath{\boxplus}}}1
    {∇}{{\ensuremath{\nabla}}}1
    {√}{{\ensuremath{\sqrt}}}1
    {⬝}{{\ensuremath{\cdot}}}1
    {•}{{\ensuremath{\cdot}}}1
    {∘}{{\ensuremath{\circ}}}1
    {⁻}{{\ensuremath{^{-}}}}1
    {▸}{{\ensuremath{\blacktriangleright}}}1
    {∧}{{\ensuremath{\wedge}}}1
    {∨}{{\ensuremath{\vee}}}1
    {¬}{{\ensuremath{\neg}}}1
    {⊢}{{\ensuremath{\vdash}}}1
    {⟨}{{\ensuremath{\langle}}}1
    {⟩}{{\ensuremath{\rangle}}}1
    {↦}{{\ensuremath{\mapsto}}}1
    {←}{{\ensuremath{\leftarrow}}}1
    {<-}{{\ensuremath{\leftarrow}}}1
    {→}{{\ensuremath{\rightarrow}}}1
    {⟶}{{\ensuremath{\longrightarrow}}}1    
    {↔}{{\ensuremath{\leftrightarrow}}}1
    {⟷}{{\ensuremath{\longleftrightarrow}}}1    
    {⇒}{{\ensuremath{\Rightarrow}}}1
    {⟹}{{\ensuremath{\Longrightarrow}}}1
    {⇐}{{\ensuremath{\Leftarrow}}}1
    {⟸}{{\ensuremath{\Longleftarrow}}}1
    {∩}{{\ensuremath{\cap}}}1
    {∪}{{\ensuremath{\cup}}}1
    {⊂}{{\ensuremath{\subseteq}}}1
    {⊆}{{\ensuremath{\subseteq}}}1
    {⊄}{{\ensuremath{\nsubseteq}}}1
    {⊈}{{\ensuremath{\nsubseteq}}}1
    {⊃}{{\ensuremath{\supseteq}}}1
    {⊇}{{\ensuremath{\supseteq}}}1
    {⊅}{{\ensuremath{\nsupseteq}}}1
    {⊉}{{\ensuremath{\nsupseteq}}}1
    {∈}{{\ensuremath{\in}}}1
    {∉}{{\ensuremath{\notin}}}1
    {∋}{{\ensuremath{\ni}}}1
    {∌}{{\ensuremath{\notni}}}1
    {∅}{{\ensuremath{\emptyset}}}1
    {∖}{{\ensuremath{\setminus}}}1
    {†}{{\ensuremath{\dag}}}1
    {ℕ}{{\ensuremath{\mathbb{N}}}}1
    {ℤ}{{\ensuremath{\mathbb{Z}}}}1
    {ℝ}{{\ensuremath{\mathbb{R}}}}1
    {ℚ}{{\ensuremath{\mathbb{Q}}}}1
    {ℂ}{{\ensuremath{\mathbb{C}}}}1
    {⌞}{{\ensuremath{\llcorner}}}1
    {⌟}{{\ensuremath{\lrcorner}}}1
    {⦃}{{\ensuremath{\{\!|}}}1
    {⦄}{{\ensuremath{|\!\}}}}1
    {‖}{{\ensuremath{\|}}}1
    {₁}{{\ensuremath{_1}}}1
    {₂}{{\ensuremath{_2}}}1
    {₃}{{\ensuremath{_3}}}1
    {₄}{{\ensuremath{_4}}}1
    {₅}{{\ensuremath{_5}}}1
    {₆}{{\ensuremath{_6}}}1
    {₇}{{\ensuremath{_7}}}1
    {₈}{{\ensuremath{_8}}}1
    {₉}{{\ensuremath{_9}}}1
    {₀}{{\ensuremath{_0}}}1
    {ᵢ}{{\ensuremath{_i}}}1
    {ⱼ}{{\ensuremath{_j}}}1
    {ₐ}{{\ensuremath{_a}}}1
    {¹}{{\ensuremath{^1}}}1
    {ₙ}{{\ensuremath{_n}}}1
    {ₘ}{{\ensuremath{_m}}}1
    {ₚ}{{\ensuremath{_p}}}1
    {↑}{{\ensuremath{\uparrow}}}1
    {↓}{{\ensuremath{\downarrow}}}1
    {...}{{\ensuremath{\ldots}}}1
    {·}{{\ensuremath{\cdot}}}1
    {Σ}{{\color{symbolcolor}\ensuremath{\Sigma}}}1
    {Π}{{\color{symbolcolor}\ensuremath{\Pi}}}1
    {∀}{{\color{symbolcolor}\ensuremath{\forall}}}1
    {∃}{{\color{symbolcolor}\ensuremath{\exists}}}1
    {λ}{{\color{symbolcolor}\ensuremath{\mathrm{\lambda}}}}1
}

\title{Proof Assistants for Teaching: a Survey}
\author{Frédéric Tran Minh \institute{Université Grenoble-Alpes, Grenoble INP, LCIS, France} \email{frederic.tran-minh@esisar.grenoble-inp.fr}
\and Laure Gonnord \institute{Université Grenoble-Alpes, Grenoble INP, LCIS, France} \email{laure.gonnord@grenoble-inp.fr}
\and Julien Narboux \institute{IRIF, Université Paris Cité, CNRS, Paris, France} \email{narboux@irif.fr}
} 

\def \titlerunning{Proof Assistants for Teaching: a Survey}
\def \authorrunning{F. Tran Minh \& L. Gonnord \& J. Narboux} 

\hypersetup{
  bookmarksnumbered,
  pdftitle    = {\titlerunning},
  pdfauthor   = {\authorrunning},
  pdfsubject  = {A survey about proof assistants for teaching},               %
  pdfkeywords = {proof assistants, teaching, interactive theorem proving, education} %
}

\definecolor{cadmiumgreen}{rgb}{0.0, 0.42, 0.24}
\definecolor{amethyst}{rgb}{0.6, 0.4, 0.8}
\definecolor{neoncarrot}{rgb}{1.0, 0.64, 0.26}
\definecolor{greyb}{rgb}{0.5, 0.5, 0.6}

\begin{document}

\maketitle

\begin{abstract}
  In parallel to the ever-growing usage of mechanized proofs in diverse areas of mathematics and computer science, proof assistants are used more and more for education.
  This paper  surveys previous work related to the use of proof assistants for (mostly undergraduate) teaching. This includes works where the authors report on their experiments using proof assistants to teach logic, mathematics or computer science, as well as designs or adaptations of proof assistants for teaching.
  We provide an overview of both tutoring systems that have been designed for teaching proof and proving, or general-purpose proof assistants that have been adapted for education, adding user interfaces and/or dedicated input or output languages.
\end{abstract}

\section*{Introduction}
\label{sec:intro}

The story of proof assistants originates in 1968 with Automath developed by De Bruijn~\cite{bruijn-de-automath-1968}. Two categories of software can be identified. 

Automatic Theorem Provers (ATP) 
take a statement as input (often expressed in a language supporting First Order Logic)  %
and act as a black box to determine if the statement is universally true or if a counter-example can be produced.
CVC-5, Z3, Vampire, Superposition are some famous examples.

On the other hand,  Interactive Theorem Provers (ITP) check for validity formal proofs designed
and written by a human user in a dedicated language, as expressive as possible
(i.e., supporting higher-order logic (HOL) or dependent type theory (DDT)). 
In ITPs, some automation is provided to help the user fill in parts of the proof considered trivial by a human being. 
The proof state (current goal and hypothesis) is displayed at any point of the proof.

Since the 1960s, numerous Interactive Theorem Provers, also called Proof Assistants, have been developed. Among the most famous still widely used nowadays are Mizar~\cite{matuszewski-mizar-2005}, Agda~\cite{goos-alf-1994,norell-towards-2007}, Coq~\cite{coquand-theorie-1985, casteran-interactive-2004}, HOL-Light~\cite{plotkin-lcf-2000} Isabelle~\cite{nipkow-isabellehol-2002} and Lean~\cite{de-moura-lean-2015}.
All these systems mainly differ in the size of their kernels,
 the underlying foundations and the expressiveness of their language.~\footnote{As the detailed description of each of these proof assistants is out of the scope of this paper, the reader can refer to~\cite{geuvers-proof-2009, nawaz-survey-2019,asperti-comparing-2003} for surveys about proof assistants.}

These ITPs have been successfully used in mathematics to formally verify proofs containing computer code (famous examples 
are proofs of the four-color theorem by  Appel and Haken in 1976, formally verified
by Gonthier in 2008~\cite{gonthier-formal-2008} and the Kepler conjecture - Hales theorem, proved by Hales in 1998
following Toth in 1953 and formally verified by the FlySpeck project in 2006 in Isabelle~\cite{hales-proof-2005,hales-formal-2015}), and also in software verification (an example is the C compiler CompCert verified by Coq in 2016~\cite{leroy-compcert-2016}).

Some work have been done to bridge the power and efficiency of ATPs with the expressiveness of ITP languages, by integrating calls to ATPs from within ITPs. Coq-hammer, SMT-Coq, Isabelle Sledgehammer, Lean-auto are some of them.

At a time when Digital Media in Education are taking center stage in innovative pedagogy, this growing role for formalization in mathematics is meeting 
with the hope of the educational community that proof technology could offer renewed approaches to teaching proof~\cite{hanna-proof-2019}. %
At the crossroads between high school and higher education, the transition to formal mathematical language
and formal proof is abrupt and many students struggle to adapt to the new expectations of their curricula~:
Hanna and Knipping review numerous research studies that have been devoted over the past 40 years to proof in education~\cite{hanna-proof-2020} and
summarize the expectations raised by proof assistants:

\begin{quotation}
(...) the use of proof technology may
present a new opportunity for fostering mathematical understanding, similar to that which has already been seen
in the use of other technology such as dynamic geometry programs. Interactive and automated theorem provers
(ITP and ATP) are designed to provide a guarantee of correctness. The question is how and to what degree
these tools can go beyond providing a guarantee and be useful in providing explanation as well. Focusing on
students’ understanding of mathematics, and of proof in particular, while making use of automated proving, will
be a promising avenue of research. Clearly, mathematics educators will have several challenging open questions
to consider when looking ahead.%
\end{quotation}

However, in a further paper, Hanna and Yan regret the low consideration for these tools in undergraduate mathematics curricula 
and lack of research evidence that they actually improve learning and understanding~\cite{hanna-opening-2021}.

In this paper, we provide a survey of the different works related to the use of proof assistants for teaching. 
Even if automated theorem provers are important tools for interactive theorem proving, and can also be used in education 
for theorem discovering or checking statements (see the line of works related to GeoGebra 
\cite{botana-automated-2015,hanna-using-2019,kovacs-geogebra-2020,Quaresma2019,Quaresma2022}), we will focus on the use \emph{interactive} theorem provers for education. 

Section~\ref{sec:teaching} exposes a variety of teaching experiences using proof assistants, from undergraduate mathematics to quasi-expert manipulation of program proofs.  Section~\ref{sec:pa:tools} describes adaptations that have been made to adapt the tools themselves, and also their user interfaces, to the special needs of education. We dedicate Section~\ref{sec:natural} to natural language interfaces.

\section{Teaching experiments using proof assistants}
\label{sec:teaching}

From the 70's, with the first reported experiments using Mizar, to now, the literature is full of experiments that somehow reflect the existing status of proof assistants: their expressiveness, of course, and also the publicity that often comes from their usage in research developments. In this section, we report a variety of teaching experiments using teaching assistants, at different levels from undergraduate to master, and targeting various topics, from logic to more advanced formal verification of programs.

\subsection{Teaching logic and meta-theory}

\label{subsec:logic}

The first experiment of teaching (propositional) logic using proof
assistants dates back to 1975, with the Mizar software. Matuszewski,
 Rudnicki and Trybulec developed an environment based on Mizar
dedicated to teaching logic~\cite{trybulec-system-1983}. This
experiment, conducted at the Warsaw University, is described in the
more recent paper~\cite{matuszewski-mizar-2005}, among other historical use of Mizar before 1989. A simplified environment (Mizar-MSE = MultiSorted with Equality), was used mostly to practice simple logical exercises, as the interface was judged austere.

Similarly, other proof assistants have been widely used in the last
40 years as computer artifacts to support teaching logic-related
topics such as propositional logic, induction, or set theory. 
The Coq Wiki~\cite{the-cocorico-coq-wiki-coq-2021,the-cocorico-coq-wiki-universities-2024} reports a list of such experiments with Coq.

Teaching logic can rely on relatively low additional machinery as proof assistants expose
natively their underlying logic. Exercising natural deduction,
sequent-calculus is thus relatively straightforward, and numerous experiments have been conducted using general purpose proof assistant or ad hoc software. 
Using proof assistants to teach the meta-theory of logic is rarer.
Some authors such as Villadsen and his co-authors have proposed using proof assistants with a deep embedding of some calculus, allowing formalization of meta-theory to teach both logic and automated deduction~\cite{villadsen-natural-2018,from-isabellehol-2020,villadsen-using-2021}.

\subsection{From logic to mathematical proofs}
\label{proofs}

Mathematical proofs are not all captured by the manipulation of logical artifacts, and the previous teaching experiments, although beneficial since they enable the formal (and quasi-syntactic) manipulation of proofs, quickly show their limit as they do not capture all subtleties of even undergraduate mathematics. Many experiments thus relate courses and even sequences of courses elaborating on a continuum of concepts from propositional logic to analysis via set theory, functions and relations, number theory, calculations and induction.

Following the initial experiment at University of Warsaw cited in Section~\ref{subsec:logic}, Retel and Zalewska
report~\cite{retel-mizar-2005} that different flavors of Mizar were then 
used, mostly at the University of Warsaw, but also at other
institutions, to  also teach axiomatic geometry (1985),
introduction to mathematics (1987), topology (c.a. 1990).
In the same paper, the authors describe their own teaching experience with
 first-year students in Bialystok learning formalization of mathematics, logic, and set theory with Mizar.

\paragraph{Experiments with Coq}

Blanc, Giacometti, Hirschowitz and Pottier proposed in 2007 a software learning environment named CoqWeb~\cite{blanc-teaching-2007}, based on Coq to teach undergraduate mathematics. %
At the University of Nice-Sophia-Antipolis, they used the software with
course pages dedicated to fields, vector spaces, polynomials over the complex-field and algebraic numbers. We describe the software in Section~\ref{sec:pa:gui}.

\paragraph{Experiments with Lean}

From 2014 to 2018, Avigad, Lewis, and Van Doorn taught an undergraduate mathematics course where proofs in natural language, symbolic logic and natural deduction, and the language of the Lean3 theorem prover were presented as three distinct languages to avoid confusion between all these activities. However, they ``emphasize that the latter two components are carried out in service of the first.''~\cite{hanna-learning-2019}. The course covered the following topics: propositional logic, elementary set theory, relations and functions, natural numbers and induction, elementary number theory, finite combinatorics, construction of real numbers, and theory of the infinite (sic), but only the first four chapters were worked out using the Lean3 prover.
At the Sorbonne University, a front-end to Lean (Deaduction~\cite{kerjean-utilisation-2022}) was also used as an executable support to teach 1st-year undergraduate maths (algebra (sets, functions), analysis (limits, continuity)).
\cite{tran-minh-use-2023} reports the development of a first-year undergraduate mathematics course at Université Grenoble-Alpes in 2022, complementing the main math class, where students used in parallel Edukera~\cite{rognier-presentation-2016} and Lean3 to explore the mathematics formal language, propositional logic, natural numbers, functions, and an overview of real analysis (numerical sequences and limits). Notably, the originality of this course is the dual navigation between two different tools.  

At Fordham University, Heather MacBeth ran a first-year undergraduate mathematics course entitled ``The Mechanics of Proof'',~\cite{macbeth-mechanics-2023} as a sequence of worked examples and exercises. Each statement is proved first with a natural language style, and then in Lean4. The course starts with very elementary ``proofs by calculation'',
and progressively introduces logic connectives (first with familiar examples, with numbers), and then natural numbers and induction, divisibility and number theory, functions, sets and relations. %

\paragraph{Experiments with various proof assistants}
In~\cite{kerjean-utilisation-2022}, the authors report on four different experiments teaching mathematics in Orsay, Montpellier, Paris, and Strasbourg using Coq, Edukera and Lean. The covered topics, all parts of first year university curriculum, range from mathematical formalism, propositional and first order logic, set theory and functions, to real numbers, numerical sequences and real functions analysis. A few students in Orsay were able to prove the intermediate value theorem and the Weierstrass theorem in Lean.

These individual experiments all rely on the fact that proof assistants provide a way to check student proofs, with various possible adaptations that we shall develop later (Section~\ref{sec:pa:tools}). 

However, to date and to our knowledge, there has been no systematic comparative study that would favor one proof assistant over another for teaching purposes. The very development of criteria for assessing the relevance of a particular software feature in a teaching context is still in its infancy.
In~\cite{bartzia-proof-2022}, Bartzia et al. studied the potential impact of different features of proof assistants.
Recently, Keenan and Omar identified criteria making a proof assistant efficient for teaching~\cite{keenan-learner-centered-2024}, in terms of feedback quality, error resilience, hint production, 
resistance to random use success, transition to pen and paper proof, possibility of testing learning definitions, learning curve, robustness, ease of installation and more.

Nevertheless, there are relatively few analyses of the impact of such new student skills on the concrete ability to understand and write classical ``pen and paper'' proofs.
Thoma et al. ~\cite{thoma-learning-2022} report on a positive effect on proof production and proof writing after using the Lean proof assistant.%
The size of the sample of students on which the study was carried out is small,
but they qualitatively conclude to a plausible link between successful proof production and student engagement with Lean, especially along two skills : the accurate use of mathematical formalism, and the hierarchical division of proofs in goals and sub-goals.
One other notable work is ~\cite{bohne-learning-2018}, in which the authors report a precise pedagogical method introducing three intermediate levels of formality to transition from a Coq proof to a textbook proof
advocating that ``the difficulty and the value of transferring the skill of proving in a proof assistant to the skill of textbook proving is extremely underestimated.''. 
They conclude on clear progress but not for all the problems related to the development of textbook proofs: at least for computer science students the formalism is not the main difficulty in learning proof, and that starting undergraduate studies with the use of ``a suitable proof assistant'' is valuable.

\subsection{Proofs assistants and program verification}

Proof assistant machinery beyond propositional logic is also taught ``for itself'': as some examples, the Coq wiki~\cite{the-cocorico-coq-wiki-universities-2024} mentions: a Theoretical Foundations of Theorem Proving Systems, by Paweł Urzyczyn, in  2006 at Warsaw University; and also a class on Automated Theorem Proving, conducted by Andrew W. Appel, in  2007 at  Princeton University. There are also classes about Coq in Strasbourg and many other universities.

More interestingly, the increasing usage of proof assistants in the
formal verification research community had a strong impact on the
evolution of graduate programs in computer science. In this particular
domain, the proofs are all done by induction on the syntax of the
programming languages which consist of many cases 
that are easier to deal with using a proof assistant, and the
research specialists have made their graduate courses evolve in the
same way as their research activities. As few examples, the proof assistant Coq has for example been used at the ENSTA french
engineering school, for a course on software safety; and at ENIEE-CNAM for a master-level class on formal methods and specification languages.

Proof assistants are here used as a way to automatize proofs
checks, or semantics specification, but not for themselves, as
advocated by Tobias Nipkow that uses HOL/Isabelle (and the
Isabelle/Isar high level language~\cite{wenzel-isabelleisar-2007} for
proof scripts) as a support of its graduate course about semantics and Hoare's
logic~\cite{kuncak-teaching-2012}.

Similarly,~\cite{feinerer-comparison-2009} reports and compares a few experiments teaching formal verification with various proof assistants such as PVS (the Prototype Verification System). Bertot has also given a course about programming language semantics using Coq~\cite{bertot:cel-01130272}.

In 2005, Delahaye reported on an experiment of teaching an introduction to programing languages semantics course with Coq to fifth-year students~\cite{delahaye-coq-2005}. The course was associated with a workshop about the design of certified software. The author asserts the students benefited from the practical nature of the course, finding it easier to start programming while receiving instant feedback than to specify and prove theorems on paper without feedback.

Perhaps the best-known material in this line of program verification and foundations of programming
languages is the series of 6 volumes entitled Software
Foundations~\cite{pierce-software-2017}, initiated by Benjamin Pierce, which
served as a basis for a software verification course he gave at
University of Pennsylvania~\cite{pierce-lambda-2009}, and many courses
after him over the world.

Finally, Coq and other proof assistants are at the heart of the courses and seminars on program reasoning, coordinated by Xavier Leroy in 2020, at Collège de France~\cite{leroy-semantics-2020}.

\section{Specializing proof assistants for education}
\label{sec:pa:tools}

The variety of teaching experiences, in terms of purpose, types of students, and the studied area, has unsurprisingly led to the observation of the necessity of a specialized proof environment. The teachers face the following difficulties: the necessity to teach a new specific language as a premise of the course itself; the proof step granularity, which often is too fine-grained for the desired effect; the difficulty of transfer to full ``pen and paper'' proof; and finally, the question of the quality of the feedback provided to students.

In this section, we review some proof assistants specifically dedicated to education (Section~\ref{sec:pa:specialized});  some interface adaptations to state-of-the-art proof assistants (Section~\ref{sec:pa:gui}); and then enumerate some tutoring features based on the improvement of feedback added to existing proof assistants (Section~\ref{sec:tutors}).

\subsection{Some proof assistants dedicated to education}

\label{sec:pa:specialized}

Teachers, or didacticians, have designed proof assistants specialized to their educational purpose. A special focus is placed on the graphical user interfaces (GUIs), on tutoring, and on simplified language interfaces. As a special case, the proof assistants dedicated to geometry form a huge part of this category.

\paragraph{Logic}

In the textbook ``A logical approach to discrete math'', by Gries and
David~\cite{gries-logical-1993}, the authors develop a mechanized
simple language to express the solutions of exercises in
logic. Realizing that students needed more feedback, Wolfram Kahl
developed in 2011 CalcCheck, a proof-checker able to verify if
\LaTeX-formatted sheets expressing proofs in this language are
correct~\cite{kahl-teaching-2011}.

Jape~\cite{bornat-japes-1996}, developed by Surin and Bornat at the University of Oxford in 1996, is a proof editor in first-order logic integrating a graphical user interface, designed to be easy to use by beginners. The interface displays lists of definitions and conjectures, it can display proofs in a boxed Fitch-style.

PhoX is a proof assistant with High-order logic (\emph{à la} HOL/Isabelle) since 2018 by Raffalli specifically to teach logic to 3rd-year and 4th-year master students~\cite{raffalli-computer-2002}. PhoX provides a graphical user interface through XEmacs, thanks to its support by ProofGeneral~\cite{goos-proof-2000}, over a limited number of tactics, designed to ease usability.

Regarding natural deduction and sequent calculus, many graphical tools enable to train with the support of an appealing graphical interface :  Panda ~\cite{gasquet-panda-2011}, The Sequent Calculus Trainer~\cite{ehle-sequent-2018}, OnlineProver~\cite{perhac-onlineprover-2024}, 
Anita~\cite{vasconcelos-anita-2023}, the Natural Deduction Planner~\cite{thompson-teaching-2017}, Yoda~\cite{machin-yoda-2011}, The Proof Tree Builder~\cite{korkut-proof-2023},  as a few.

\begin{figure}[!htb]

\begin{subfigure}{0.5\textwidth}
\begin{minipage}{0.99\linewidth}
 \vspace*{-0cm}
\includegraphics[width=0.99\linewidth]{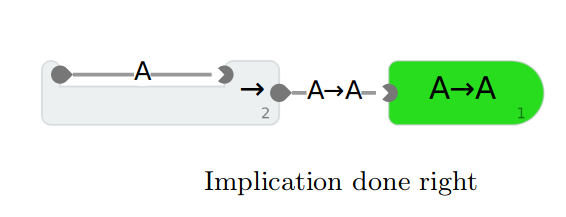}   \\
\medskip \\
\medskip \\
\includegraphics[width=0.85\linewidth]{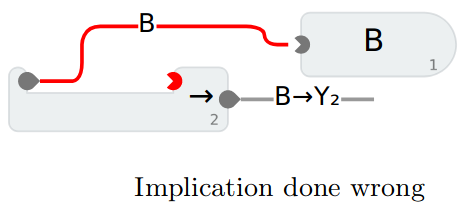}  
\end{minipage}
\caption{The incredible proof machine (Breitner).}
\label{fig:logicbreitner}
\end{subfigure}
\begin{subfigure}{0.5\textwidth}
\includegraphics[width=0.85\linewidth]{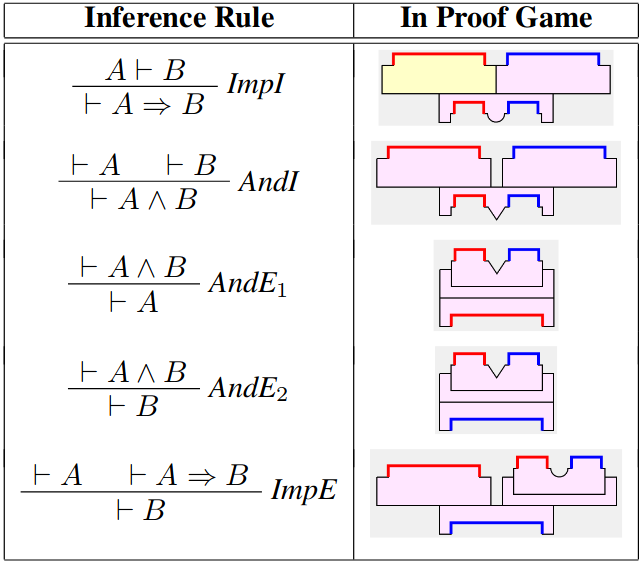} 
\caption{Polymorphic blocks in Logic Puzzle (Lerner).}
\label{fig:logiclerner}
\end{subfigure}

\caption{Logic games}
\label{fig:logicgames}
\end{figure}

There also exists some educational point-and-click environments aimed at solving logic puzzles, in propositional and first-order logic: among them, we can cite \emph{The incredible proof machine} by Breitner~\cite{blanchette-visual-2016}, 
the \emph{Logic Puzzle} proof game of Lerner~\cite{lerner-polymorphic-2015}, QED by Terence Tao~\cite{tao-qed-2018} or Carnap by Leach-Krous~\cite{leach-krouse-carnap-2018}. %
Figure~\ref{fig:logicgames} depicts some of these interfaces, illustrating how they can make learning logic visual and playful.

\paragraph{Undergraduate mathematical proofs}

PhoX~\cite{raffalli-computer-2002}'s environment also comes with some analysis topics (limits continuity, intermediate value theorem),  topology (properties of closures, of connected sets), and algebra and number theory (symmetric group, rings, prime numbers), thought for undergraduate students.

Lurch~\cite{carter-lurch-2013} is described by its authors, Carter and Monks, as a word processor enabling to typeset mathematics,
equipped with verification features designed for student use. In a graphical environment, the user enters a mathematical text in natural language,
but selects specific expressions to mark them as ``meaningful''. Lurch verifies the logical coherence of meaningful expressions and provides feedback, 
and ignores the rest of the text. Three levels of input constraints and automation are available: formal proofs, semiformal proofs, and expository proofs.
The teacher can prepare a dedicated rule set (logic system, axioms). The system can act as a tutor, grading, coaching and providing hints.  Two experiments using Lurch with students were conducted~\cite{carter-formal-2016}: in an introduction to formal logic and mathematics courses; the topics covered were logic, set theory, functions, algebra, inequalities, relations, number theory, and combinatorics.

Diproche (“DIdactical PROof CHEcking”), an educational variant of Naproche~\cite{kuhlwein-naproche-2009},  checks student solutions to proving exercises written in a controlled natural language~\cite{carl-natural-2022}.
Its language is adapted to specific areas common in introductory courses to proof method, and supports 
exercises in propositional logic, set theory, functions and relations, number theory, axiomatic geometry, and elementary group theory.
An ATP is used to provide feedback to check inference steps, detect typical error patterns and provide counterexamples to false claims.
The modules for boolean set theory and elementary number theory of the system were used in 2020/2021 in a first-semester course at the University of Flensburg.

Edukera has been developed in 2016 by Rognier and Duhamel~\cite{rognier-presentation-2016}. It appears as a web interface: the user interacts with the software by pointing and clicking, there is no text-based language to typeset. A database of 900 exercises is provided, including 200 formalization exercises.
In formalization exercises, the student is proposed a sentence in French or English (like ``Someone read all books by Victor Hugo''), and is asked to formalize it using quantifiers, variables and given predicates; to do so, the user has to click on connectives in a prefix order. The resulting answer is dynamically displayed. In proof exercises, a statement is to be proven. To show the goal, one must select suitable rules to apply 
(introduction, elimination, apply lemmas, \ldots); the current state of the proof can be displayed in Fitch's style or Gentzen's sequent-calculus style. %

\paragraph{The special case of geometry }
\label{subsec:geom}

\begin{figure}[!htbp]

\begin{subfigure}{0.5\textwidth}
\hspace*{+0.4cm}\includegraphics[width=0.77\linewidth]{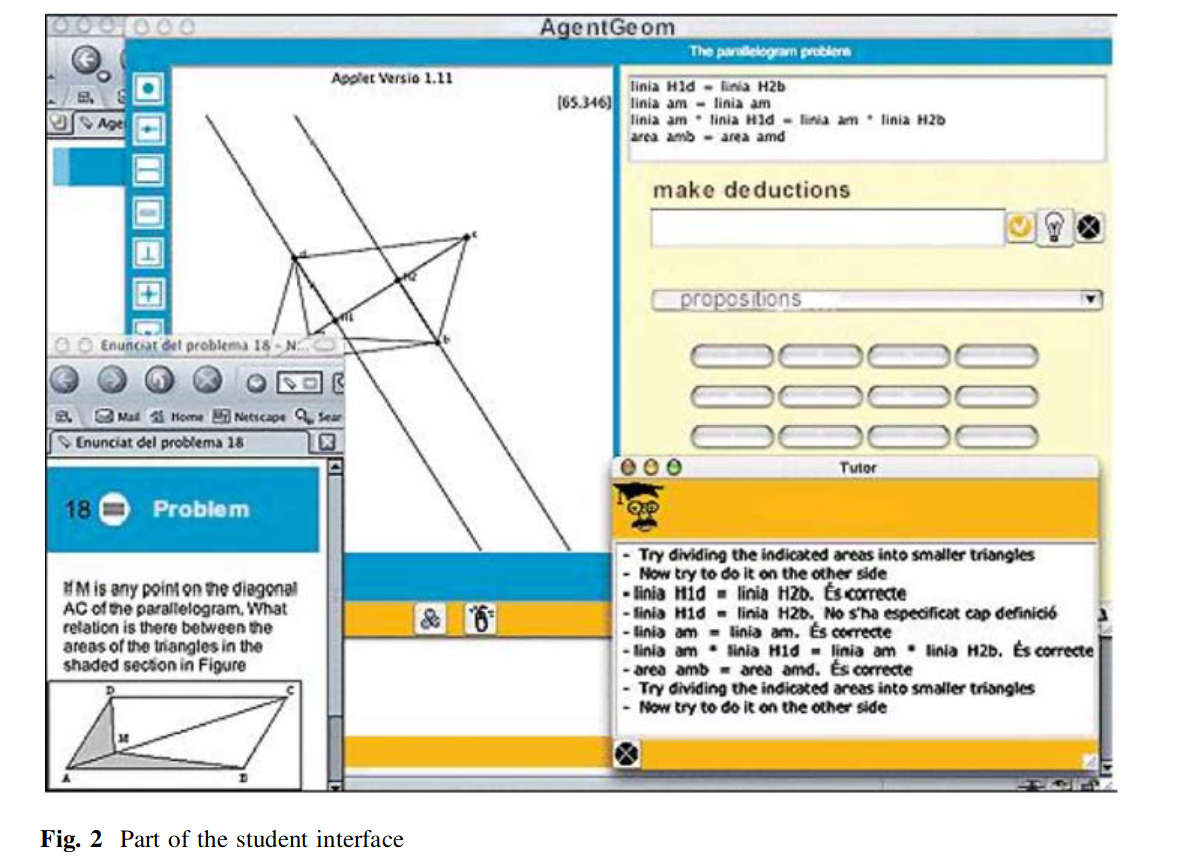}
\caption{Agent Geom}
\label{fig:geometryagentgeom}
\end{subfigure}
\begin{subfigure}{0.5\textwidth}
\hspace*{+1cm}\includegraphics[width=0.84\linewidth]{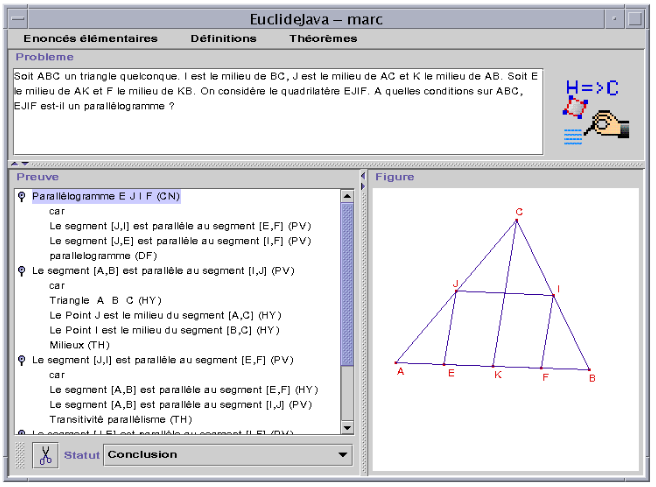}
\caption{Baghera}
\label{fig:geometrybaguera}
\end{subfigure} \\
\begin{subfigure}{0.5\textwidth}
\hspace*{-1cm}\includegraphics[width=1.09\linewidth]{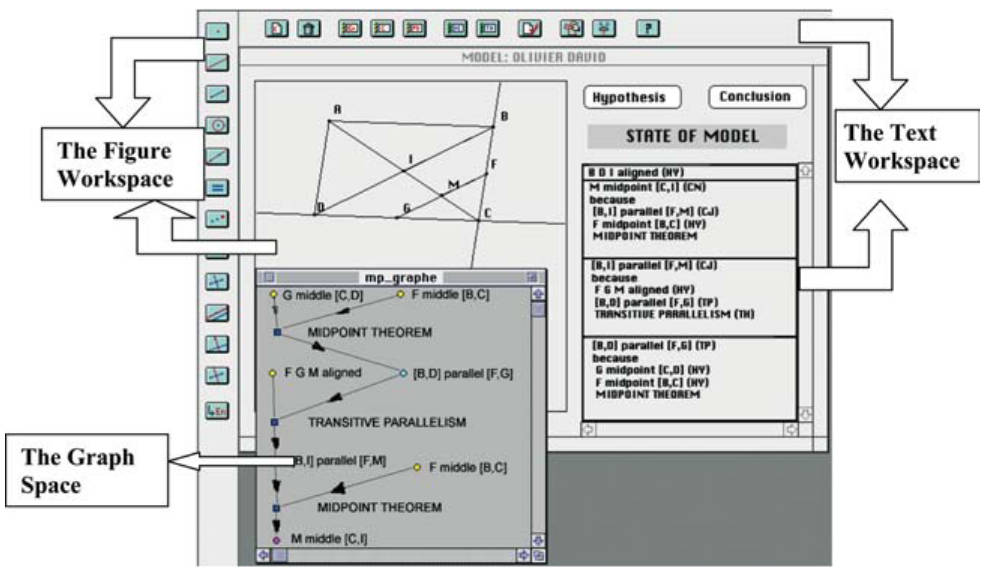}
\caption{Cabri-Euclide}
\label{fig:geometrycarbrieuclide}
\end{subfigure}
\begin{subfigure}{0.5\textwidth}
\hspace*{+1.3cm}\includegraphics[width=0.79\linewidth]{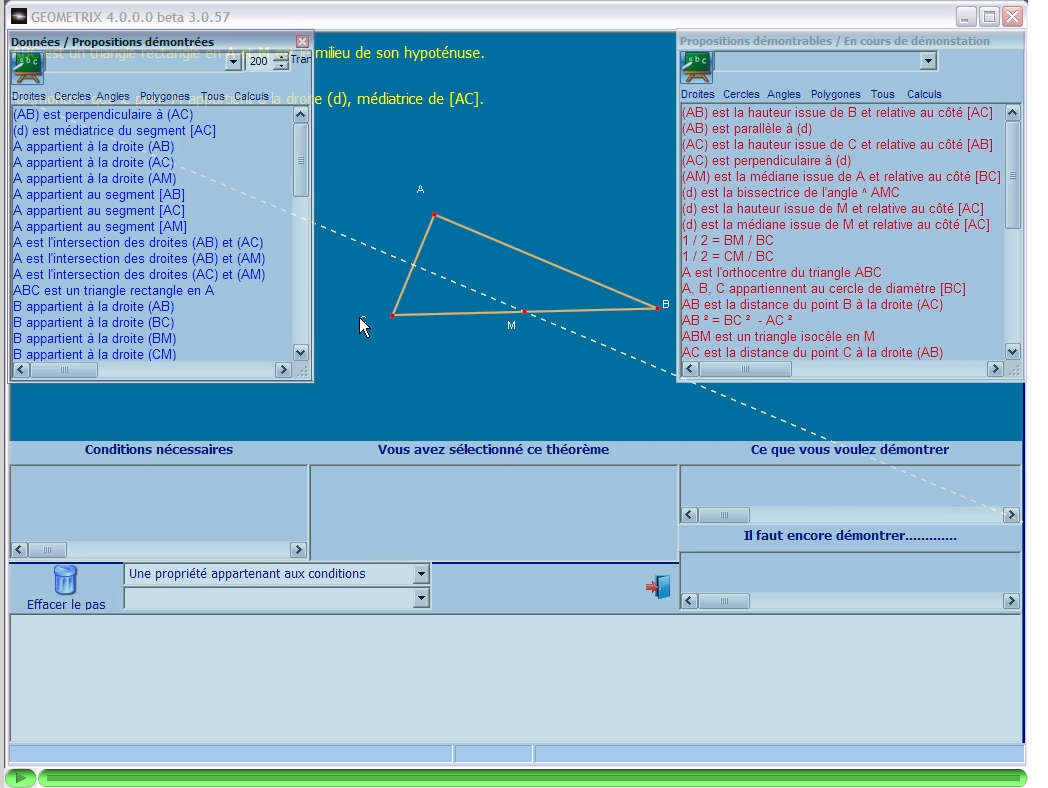}
\caption{Geometrix}
\label{fig:geometryGeometrix}
\end{subfigure} \\
\begin{subfigure}{0.5\textwidth}
\hspace*{+0.5cm}\includegraphics[width=0.74\linewidth]{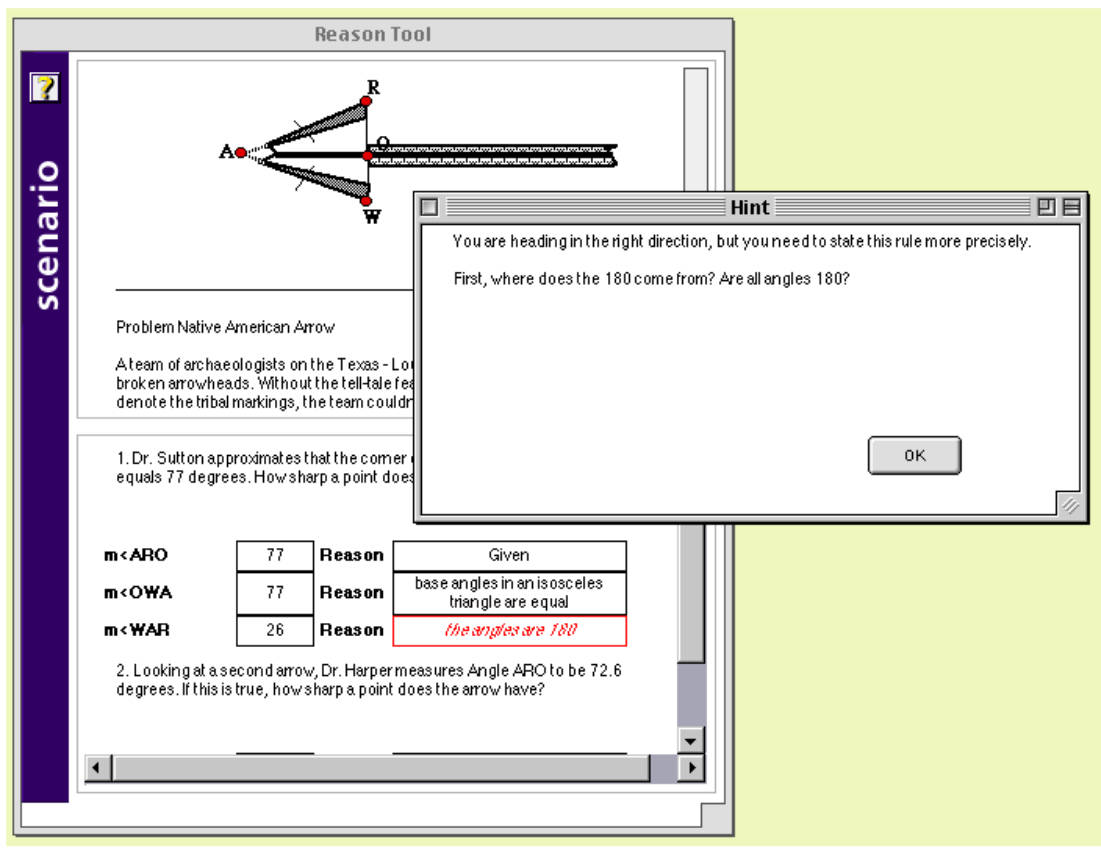}
\caption{Geometry Explanation Tutor}
\label{fig:geometryGeometryExplanationTutor}
\end{subfigure}
\begin{subfigure}{0.5\textwidth}
\hspace*{+1.1cm}\includegraphics[width=0.84\linewidth]{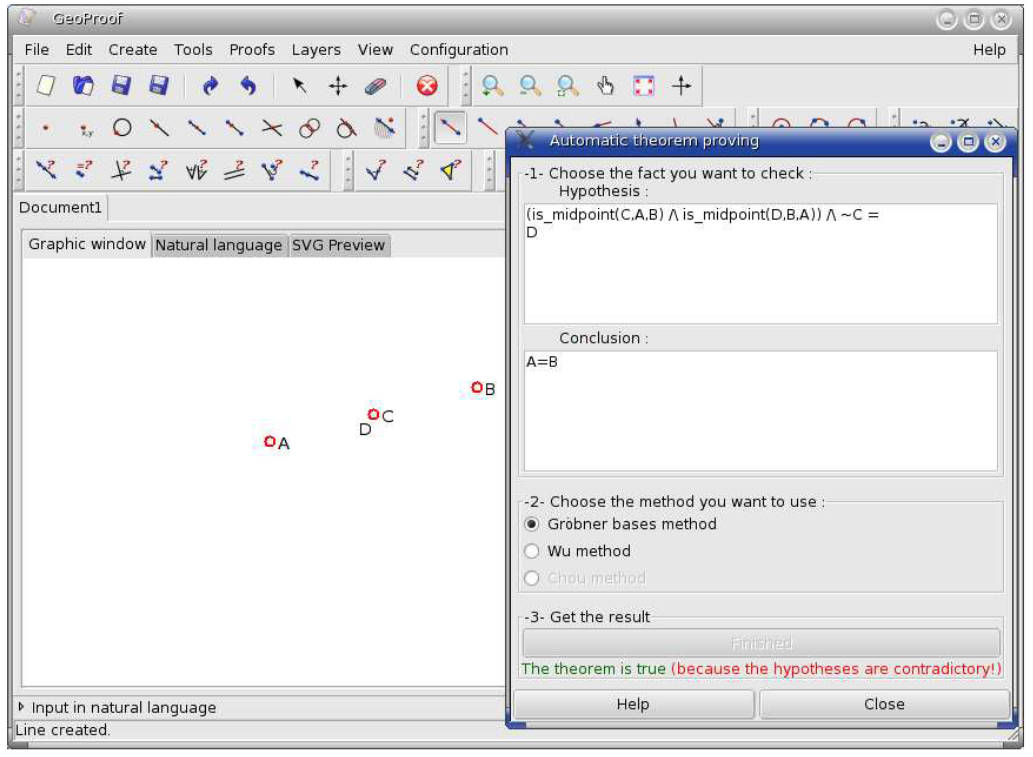}
\caption{GeoProof}
\label{fig:geometryGeoProof}
\end{subfigure} \\
\begin{subfigure}{0.5\textwidth}
\includegraphics[width=0.89\linewidth]{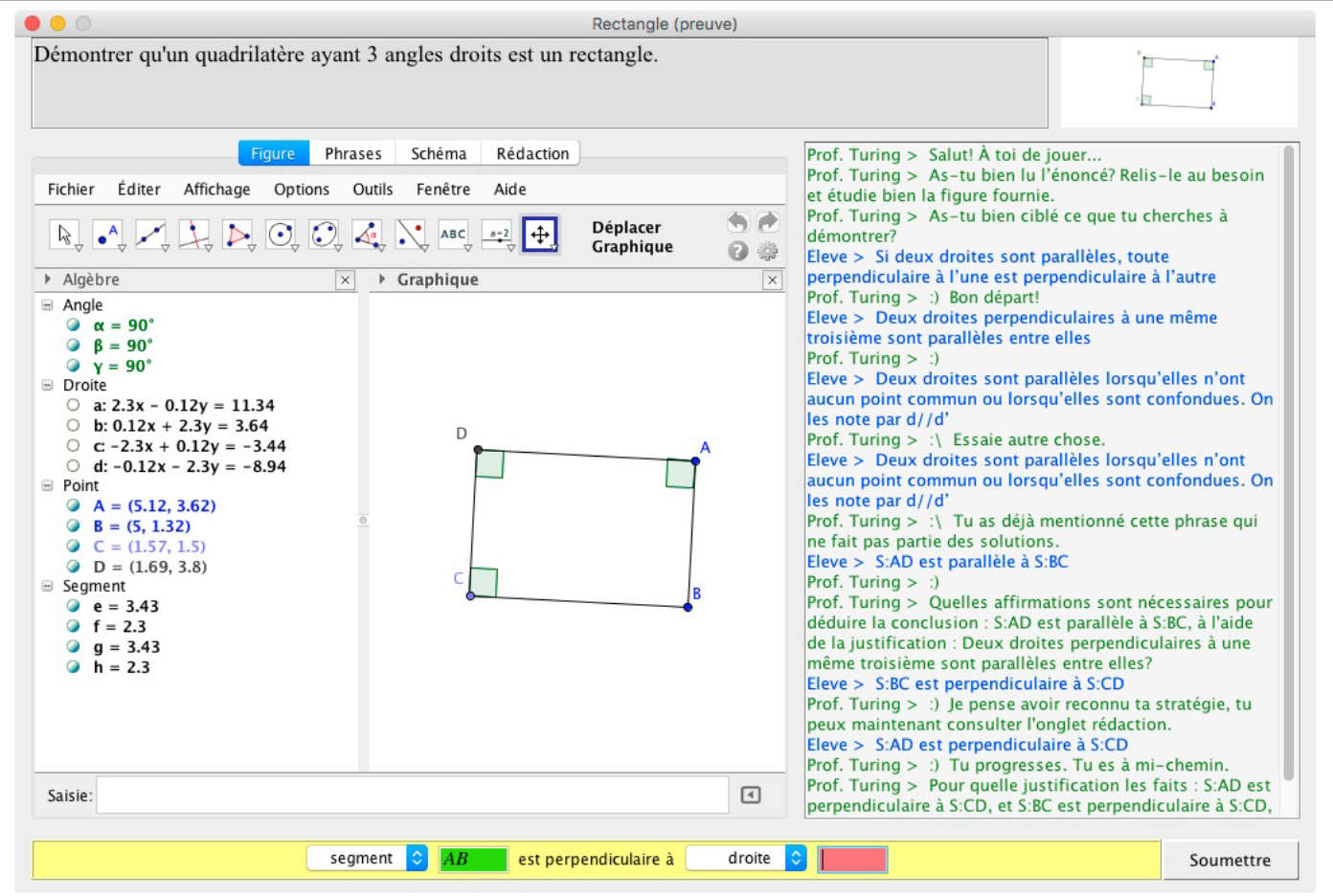}
\caption{QED-Tutrix}
\label{fig:geometryQedTutrix}
\end{subfigure}
\begin{subfigure}{0.4\textwidth}
\hspace*{+1.3cm}\includegraphics[width=0.99\linewidth]{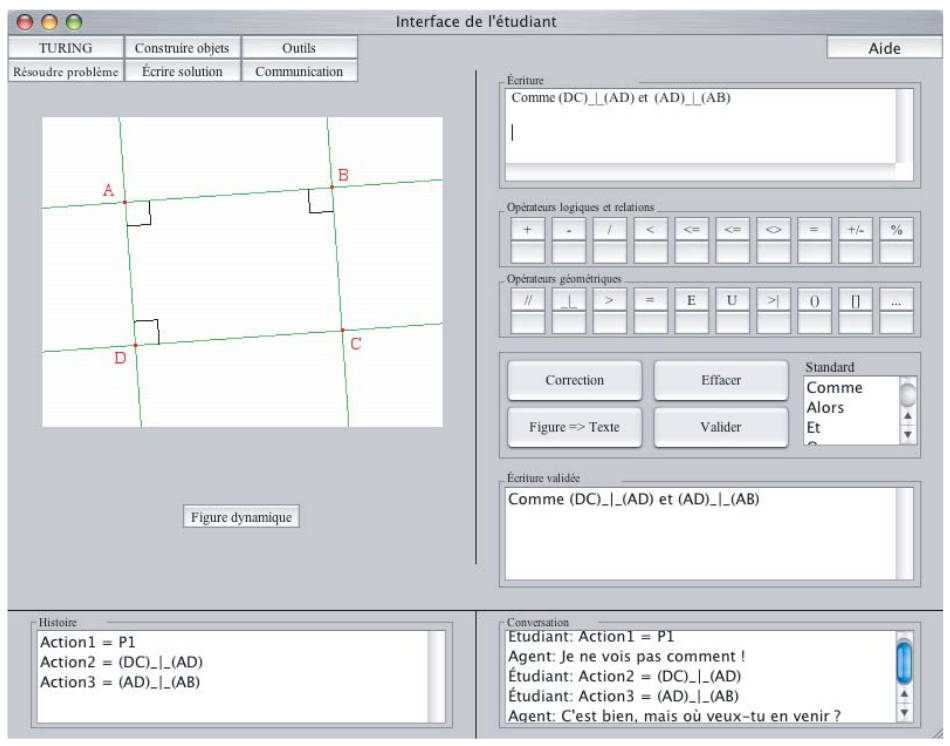}
\caption{Turing}
\label{fig:geometryTuring}
\end{subfigure}

\caption{Geometry software featuring not only proof verification, but also tutoring.}
\label{fig:geometry}
\end{figure}

Several e-learning environments have also emerged to support proof learning in the context of synthetic geometry in secondary education. 
They differ from general-purpose proof assistants for two reasons. First, the feedback and error messages are designed for use in the classroom. 
Second, the systems are often not of general purpose from a logical point of view: there are proper proofs that can not be formalized in these systems (using analytic geometry, using complex numbers,\ldots), 
the user can not introduce its own lemmas or definitions.
These geometric systems do not deal with degenerated cases and hence can present proofs without case distinctions as a list of modus-ponens steps along with figures displayed with embedded dynamic geometry software. They usually do not allow reasoning by contradiction. There are many software packages of this type, such as AgentGeom~\cite{cobo-agentgeom-2007}, Baghera~\cite{webber-baghera-2001},  Chypre~\cite{bernat-chypre:-1993},  Cabri Euclide~\cite{luengo-cabri-euclide-1998}, Geometrix~\cite{gressier-geometrix-2006}, Geometry Tutor~\cite{anderson-geometry-1985}, Geometry Explanation Tutor~\cite {aleven-towards-2001}, Mentoniezh~\cite{py-geometry-1993}, QED-Tutrix~\cite{leduc-qed-tutrix-2016},   Turing~\cite{richard-amelioration-2007}, some of which are illustrated in Figure~\ref{fig:geometry}. 
 It is not possible to give an overview of these systems here. 
They differ in many aspects such as the feedback they provide to students' errors and how they can provide hints by computing the distance from a pre-defined list of solutions or using key steps provided by the teachers, the way proofs are stored behind the scenes, if they separate the exploration and problem-solving phase of the exercises, \ldots
 For a survey (in French) comparing these systems from the didactic point of view see~\cite{tessier-baillargeon-etude-2017}.
 There are also many tools incorporating automatic deduction engines inside dynamic geometry software, such as GeoProof~\cite{narboux-graphical-2007}, GCLC~\cite{janivcic2006gclc} and Geogebra ART~\cite{botana-automated-2015}, which can be used to verify statements.
GeoGebra is also able to discover some statements using automated deduction techniques or interactively with the student using locus computations~\cite{Quaresma2019,hanna-using-2019,kovacs-geogebra-2020,kovacs2021}.

\subsection{Development of user interfaces on top of proof assistants}
\label{sec:pa:gui}

User interfaces make sense particularly for teaching purposes, as they partly respond to essential challenges: interactivity and visualization are essential to understand the structure of proofs, and provide feedbacks; and graphical helpers could help to manipulate the underlying sometimes complex language.

As programming languages, proof assistants are ``naturally'' equipped with graphical user interfaces and development environments. In addition to editing facilities, the ability to update the proof state was their minimal interactivity requirement. Individual proof assistants often have their own graphical user interface, like CoqIDE for Coq, XIsabelle~\cite{cant-xlsabelle-1995} or Isabelle/jEdit for Isabelle. Web interfaces (jsCoq~\cite{gallego-arias-jscoq-2017} for Coq,  Clide~\cite{hutchison-web-2013} for Isabelle, The Lean 4 web editor~\cite{nawrocki-lean-2022} for Lean) have the advantage of facilitating maintenance or adoption to large classes. Other web interfaces include  Alfie~\cite{the-programming-logic-group-in-goteborg-implementation-nodate},  CoqWeb~\cite{blanc-teaching-2007}, %
ProofWeb~\cite{wiedijk-teaching-2007}, Trylogic~\cite{terrematte-trylogic-2015}, PeaCoq~\cite{robert-introducing-2014}.

Among all these interfaces, the Emacs-based front-end ProofGeneral is the most generic front-end for interactive assistants,
featuring toolbar and menus, proof tree visualization,
script edition and management, and proof by pointing; which supports
more than ten proof assistants, including Coq, Isabelle, PhoX, HOL Light and others. The advent of Visual Studio also enabled the diffusion of provers plugins, with enhanced capabilities. As plugins become more pervasive, some APIs (Lean Proofwidgets~\cite{nawrocki-extensible-2023}, discussed below) have been designed for participative evolution of interfaces for the most common provers.

\begin{figure}[!htbp]

\begin{subfigure}{0.5\textwidth}
\includegraphics[width=0.64\linewidth]{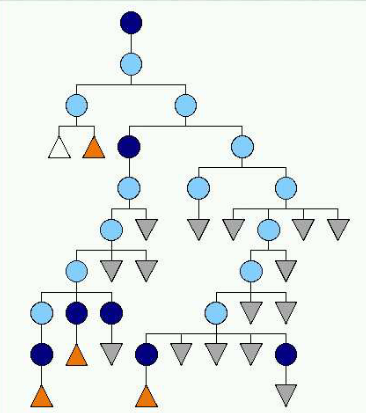}
\caption{Omega L$\Omega$UI.}
\label{fig:displayproofomega}
\end{subfigure}
\begin{subfigure}{0.5\textwidth}
\includegraphics[width=0.84\linewidth]{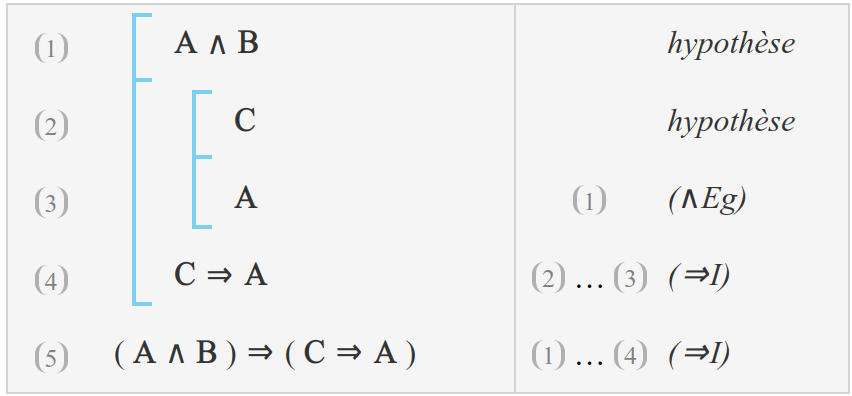}
\caption{Edukera.}
\label{fig:displayproofedukera}
\end{subfigure} \\
\begin{subfigure}{0.5\textwidth}
\includegraphics[width=0.99\linewidth]{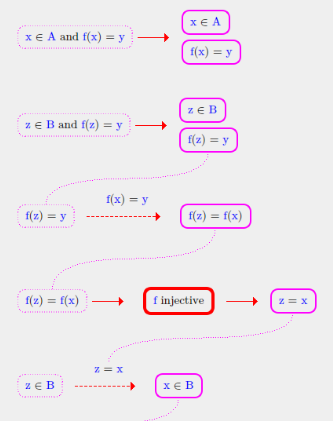}
\caption{D$\exists\forall$duction.}
\label{fig:displayproofdeaduction}
\end{subfigure}
\begin{subfigure}{0.5\textwidth}
\includegraphics[width=0.99\linewidth]{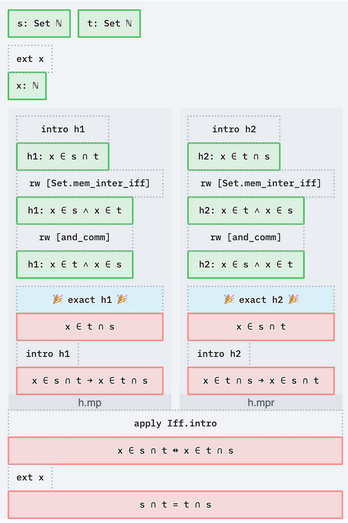}
\caption{Paperproof.}
\label{fig:displayproofpaperproof}
\end{subfigure}

\caption{Proof visualizations.}
\label{fig:proofvisualizations}
\end{figure}

An additional and required feature for education is the capability to provide a visualization of the proof itself: displaying proof trees, Fitch or Genzen style while manipulating these proof techniques. Figure \ref{fig:proofvisualizations} show a few examples. Our oldest reference is CPT~\cite{scheines-computer-1994}, in which ``Goal Tree display'' was designed with ergonomy in
mind when proceeding forward and backward proofs.
Various proof displays are available for instance in Proofweb,
Deaduction~\cite{kerjean-utilisation-2022} and Paperproof~\cite{karunus-lean-2024}. 
Paperproof is a Visual Studio code
extension for Lean that displays a boxed proof tree-like representation of the current active -~possibly incomplete~- proof.
With a clever use of nested boxes, colors and curved arrows, it manages to render backward chaining as well as forward chaining and variable scopes.

Among proofs, interfaces have been developed to help visualize the objects ``under proof'' themselves. Geoview, an extension of PCoq designed for geometry~\cite{bertot-visualizing-2004} and GeoProof~\cite{narboux-graphical-2007} for instance, displays geometric figures. The ProofWidgets' extension to Lean~\cite{nawrocki-extensible-2023} supplied examples include graphical presentations like commutative diagrams, Venn diagrams, geometry diagrams, animated plots of parametric functions, and graph representations.

On the proof presentation side, JsCoq and Waterproof provide a ``mixed document'' feature that allows the formal proof source code to be embedded into formatted HTML or LaTeX content.

Bertot, Thery and Kahn conceived in 1994 a ``proof by pointing'' algorithm~\cite{bertot-proof-1994}
 which associates the backward application of an inference rule and a residual sub-expression to each possible user selection of a sub-expression of a goal sequent. 
 The implementation results in a graphical interface that enables the user to direct her proof by selecting sub-expressions, 
 the tactic corresponding to the inferred rule being automatically applied subsequently to the mouse selection
 The software has been implemented in Coq and for other proof assistants (Isabelle, HOL). 
 These principles were later embedded in a wider multiprocess interface first called CtCoq~\cite{bertot-ctcoq-1999}, implemented in Lisp ;
 PCoq  (2001-2003) is a reimplementation in Java~\cite{amerkad-mathematics-2001}.

\begin{figure}[!htbp]

\begin{subfigure}{0.5\textwidth}
\includegraphics[width=0.99\linewidth]{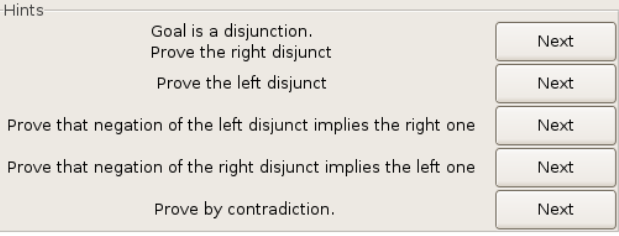}
\caption{Hints to prove a disjunction in Papuq.}
\label{fig:proofpointingpapuq}
\end{subfigure}
\begin{subfigure}{0.5\textwidth}
\includegraphics[width=0.99\linewidth]{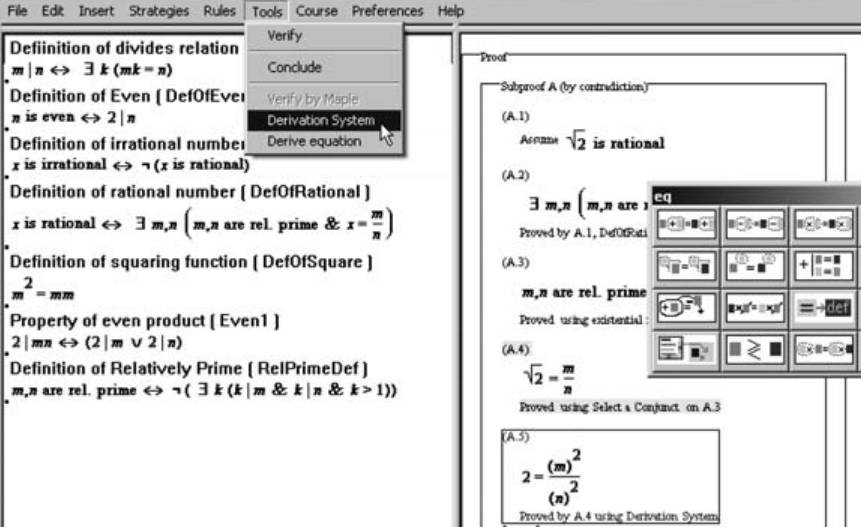}
\caption{TPE/Epgy user interface.}
\label{fig:proofpointingepgy}
\end{subfigure} \\
\begin{subfigure}{0.5\textwidth}
\begin{minipage}{0.99\linewidth}
 \vspace*{-0cm}
\includegraphics[width=0.49\linewidth]{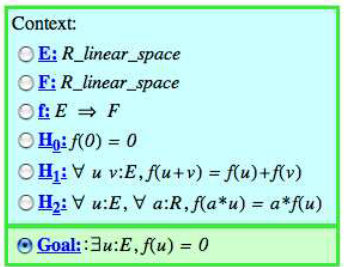}   \\
\includegraphics[width=0.99\linewidth]{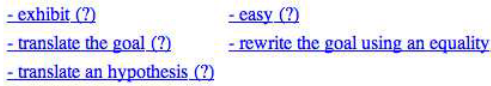}  
\end{minipage}
\caption{Proving that a linear map preserves vector space structure in Coqweb.}
\label{fig:proofpointingcoqweb}
\end{subfigure}
\begin{subfigure}{0.5\textwidth}
\includegraphics[width=0.99\linewidth]{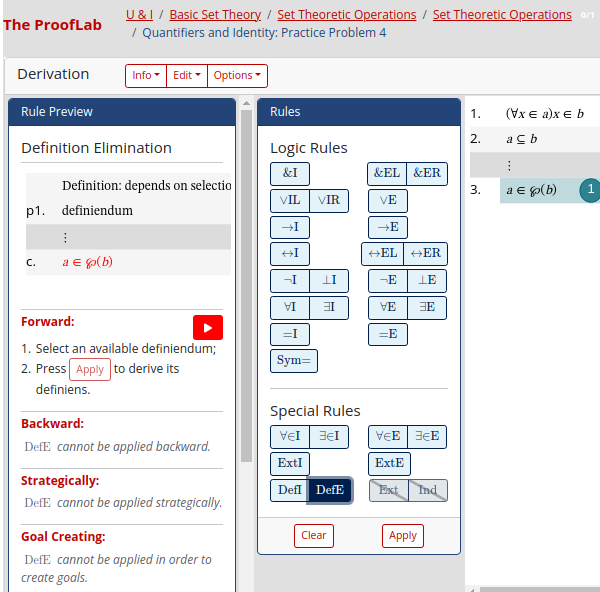}
\caption{Proof lab user interface.}
\label{fig:proofpointingprooflab}
\end{subfigure} \\
\begin{subfigure}{0.5\textwidth}
\includegraphics[width=0.99\linewidth]{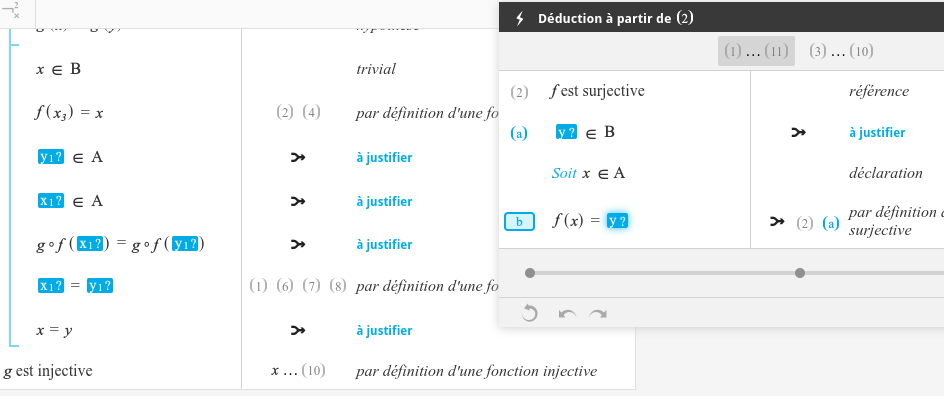}
\caption{Point'n click proof in Edukera.}
\label{fig:proofpointingedukera}
\end{subfigure}
\begin{subfigure}{0.5\textwidth}
\includegraphics[width=0.99\linewidth]{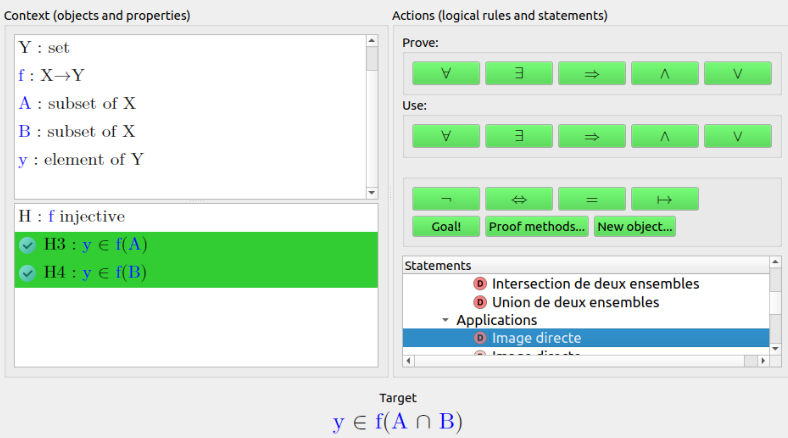}
\caption{A proof that injective applications preserve intersection in Deaduction.}
\label{fig:proofpointingdeaduction}
\end{subfigure} 
\caption{Proof by pointing.}
\label{fig:proofpointing}
\end{figure}

This concept of ``proof by pointing'' has been used in a variety of proof assistants, used or not for educational purposes: CtCoq, %
Papuq~\cite{chrzaszcz-papuq-2007}, CoqWeb, Edukera, Deaduction, Waterproof,
ProveEasy~\cite{burstall-proveeasy-2000}, for instance, proposes such a feature,
as illustrated in Figure~\ref{fig:proofpointing}.

Some extensions have been specially designed to target teaching activities.
Some are rather technical yet important: ProofWeb,  Trylogic, ProofLab/AProS~\cite{sieg-apros-2007}, Edukera~\cite{rognier-presentation-2016},  as a few, provide access to exercise database that can be enriched by the teacher; and/or link to existing classical material to understand a particular proof technique.
Trylogic can additionnaly generate automatically new conjectures to prove or refute.
Deaduction's database contains exercises in logic, set theory, functions and relations, and basic analysis, as needed for a first-year undergraduate course. The Theorem Proving Environment (Epgy project)~\cite{sommer-proof-2004}, ProofWeb, and Trylogic provide the ability for the teacher to replay student proofs or track student progression~; Trylogic even interfaces with Moodle.
Finally, ProofBuddy~\cite{karsten-proofbuddy-2023} (a web interface for Isabelle) collects ``fine-grained'' data about the way students interact with the Isabelle proof assistant.

\paragraph{Theory adaptations, simplifications}

Papuq~\cite{chrzaszcz-papuq-2007} is an extension of CoqIde designed for teaching logic and set theory to first-year students at Warsaw University. The authors
first propose to expose only the aspects of Coq type theory that are relevant to teach to students, and slightly adapt its vocabulary. They discuss the presentation of sets as predicates, partial functions, relations and functions as primitive objects and functional relations.
 They call these adaptations ``Naïve Type Theory'' after Constable~\cite{schwichtenberg-naive-2002} and Kozubek~\cite{miculan-search-2008}.

The formalization and manipulation of functions over a subset of a type universe (typically: a subset of $\mathbb{R}$) 
or undefined terms (e.g., undefined division result, limit, derivative, supremum)
in a proof assistant is known to often be an additional burden irrelevant to the study of undergraduate mathematics.
Several authors tried to tackle this difficulty in a teaching context. Sommer and Nuckols proposed~\cite{sommer-proof-2004} to handle
 functions and relations defined on a partial domain (e.g.,$x \mapsto 1/x$) by generating side conditions either automatically justified or returned to the user as ``proof obligations'' to be proved, the management of the proof obligations being integrated into the internal system of inference rules.
 The method has been implemented in the Theorem Prover Environment (Epgy project). Coen and Zoli adopt a theoretical modelization
 of undefined terms by elements of a partial setoid over an asymmetric relation of ``quasi equality''~\cite{coen-note-2007}. They implemented their idea in
 the Matita theorem prover~\cite{bjorner-matita-2011}.

The CoqWeb interface for Coq was designed to target teaching activities. It features proof by pointing and clicking like CtCoq,
but does not, on purpose, infer the right tactic from a pointed expression, since ``it was too much an invitation to click without relating it to any concept''~\cite{blanc-teaching-2007}.
Instead, the student can choose from a list of possible moves given in informal natural language, each of which, when clicked, triggers a specific Coq tactic.
The original available Coq tactic library has been restricted to the minimum necessary for educational purposes, and some of them have been adapted to provide modes: in student-guided mode, a hint is provided for the next move; in teacher mode, expert teachers can declare definitions, axioms and exercises.

Other tools (Waterproof, ProveEasy, TPE/Epgy, WinKE~\cite{dagostino-winke-1998}, ETPS)
also provide a teacher mode to provide additional (simplifier) rules and
axioms.
Finally, $\Omega$mega Tutor, TPE/Epgy and Tutch provide automation to adapt proof step granularity.

\subsection {Improving the feedback}
\label{sec:tutors}

A major advantage of proof assistants is that they provide immediate feedback on the validity of a proof step. Error messages returned by the assistant are generally not usable by students: the feedback is therefore binary. A crucial issue if proof assistants are to be used in teaching, is that they should be able to provide feedback that sheds light on the error made.
Below are some responses to this problem.

The proof tutor Tutch of Abel, Chang and Pfenning consists of a proof language and a command line
proof checker program which can annotate proof steps; in case of erroneous or insufficiently justified steps, the program provides error messages, and assumes them to continue checking to the end of the input.  Terms of the proof language are structured by bracket-surrounded frames corresponding to the scope of introduced symbols or hypotheses. Several flavors (from proof-term style to assertion level via declarative style) of the language are available, depending on the teaching activity purpose.

Inspired by the results of the Dialog Project (2001)~\cite{benzmuller-tutorial-2003}, Dietrich, Benzmüller and Schiller integrated a tutoring module to the $\Omega$mega proof assistant~\cite{benzmuller-mega-1997,schiller-proof-2008}. The system invites the student user to a dialog in natural language where she submits her next proof step. The system replies to the student proposition along three criteria: correctness, proof granularity, or relevance. In case of too coarse-grained step, the user is prompted for an additional justification.

The idea of introducing ``buggy rules'' into a problem solver to model bugs in algorithms or fallacies in students' reasoning seems to originate
in the late 1970s with Brown, Burton and Larkin~\cite{brown-representing-1977,brown-repair-1980}. The authors conceived a system for synthesizing
a model of pupil misconceptions in school basic arithmetics, and applied it by designing for student teachers a game simulating the possible errors of a pupil~\cite{brown-diagnostic-1978}. Following the same idea, Farrell et al., designed a tutor to teach how to program in Lisp~\cite{farrell-1984-1984}.
This work later inspired Zinn in the 2000s when developing Slopert aimed at diagnosing errors in symbolic differentiation.
Recently, Carl named anti-ATP~\cite{carl-using-2020} the idea, implemented in Diproche, of %
trying to verify students' erroneous proof steps with an ATP with false rules or axioms as input.
Different types of false rules (logical fallacies, axioms of erroneous distributivity, commutativity, monotonicity, or rules of false analogy), 
based on teaching experience, lead to the prediction of different possible types of mistakes, thus providing precise feedback to the students.

An expected feature when it comes to put a proof assistant in students' hands is the ability to provide clues about the next proof step.
Most software programs that implement ``proof by pointing'' are equipped with at least a basic hint production feature : as mentioned above, CtCoq, Papuq, Coqweb, Lurch, Edukera, Waterproof, LeanVerbose, as well as ETPS~\cite{andrews-etps-2004} or CPT, as a few, are able provide minimal strategic help, such as detecting the main connector of the goal and proposing the correct way of introducing it.
To go beyond that, some tutoring software rely on a concept of proof planning~\cite{lusk-use-1988}. X-Barnacle~\cite{lowe-use-1997} is
a graphical interface based on the automated prover CLAM, displaying a proposed proof tree for a given statement, in which the user can intervene to modify and reorient the proof. With $\Omega$-Ants, Benzmuller and Sorge invented a multi-agent mechanism~\cite{carbonell-blackboard-1998} to
compute suggestions on how to further construct a proof.

\section{Proof output or input in natural language}
\label{sec:natural}

Proof assistants have often been criticized for the fact that their language, which is essentially a computer language, is too far away from the vernacular of mathematics, and that it requires an excessive depth of detail in the justifications, masking the key ideas of proofs.
These peculiarities put off both the mathematical community, which was reluctant to commit to formalization, and the teaching community, which saw the use of proof assistants as ineffective in transmitting the language habits of mathematicians.

Based on these observations, many authors have sought to bring the language of proof assistants closer to the natural language of mathematics,
in two main ways :

\begin{itemize}
 \item  Using controlled natural languages (CNL)~\cite{kuhn-survey-2014} as input. These are languages that have the rigor and precision of a programming language, but whose vocabulary,  
 syntactic structure, and tolerance for ellipses and missing steps allow them to resemble natural language; 
 their interpretation generally requires a high level of automation.

 \item Conversely, converters have been developed to convert a proof script in the native language of a proof assistant into a "human-readable" language
 emulating the vernacular language. The presentation of the output can be formatted, for example in \LaTeX. The structured nature of the proof term 
and the use of interactive formats (html, javascript) allows a level of detail to be selected by the user.

\end{itemize}

\subsection{Translating formal proof into natural language}
 
 $\Omega$mega, implemented in Lisp by Benzmüller~\cite{benzmuller-mega-1997} is based on a concept of ``proof planner'' which consists in identifying 
 different levels of granularity in the proof, using forward and backward state-space search. It supports ATP integration (Otter) and embeds a computer algebra system (CAS) to help extract proof plans. It relies on Proverb~\cite{huang-presenting-2000} to output proofs in a natural language at a user-selectable level of abstraction.

  Theorema~\cite{buchberger-theorema-2006,windsteiger-theorema-2013}  is an environment been developed since 1995 by Buchberger and then by Windsteiger on top of Mathematica (and in the language of Mathematica scripts), which aims at associating theorem proving with computing and solving facilities. It generates proofs for statements expressed in untyped higher-order logic, in a style that imitates natural language, but seemingly still supports only a few theories (eg predicate logic, induction on natural numbers).

 In 2013, Ganesalingam and Gowers wrote a program that proves elementary undergraduate-level statements
 trying to produce a ``human-style output''~\cite{ganesalingam-fully-2013}; to do so they try to analyze and reproduce typical stereotypes
 mathematicians apply when faced with these kinds of problems. The outputs of the program were mixed with human solutions written by an undergraduate student and a  PhD student, and submitted to the internet community to guess which of them were auto-generated: according to the authors, the results were encouraging in that 
 ``the program did reasonably well at fooling people that it was human''.

In a recent Workshop~\cite{massot-formal-2023}, Patrick Massot presented a tool to convert a Lean proof into a natural language proof. The output format is an interactive HTML document and allows to select the desired granularity by expanding or collapsing any argument at any level in the proof.~\footnote{A demonstration is visible at \url{https://www.imo.universite-paris-saclay.fr/~patrick.massot/Examples/ContinuousFrom.html}}
 
Figure \ref{fig:naturallanguagesoutput}  demonstrates the output languages imagined by Benzmüller,  Ganesalingam and Gowers and Massot look like.

\newcommand{\pow}{\^{}}

\begin{figure}[!tb]

\begin{subfigure}{0.99\textwidth}
\fbox{\begin{minipage}{0.99\linewidth}
\scriptsize
\textbf{Theorem:} \textit{Let there be a y in Z such that there exists a z in Z such that x*y=z
and there is no d in Z such that d is a common divisor of y and z for all x in Q.
Therefore sqrt(2) isn't rational.}

\textbf{Proof:}
Let there be a y in Z such that there exists a z in Z such that x*y=z
and there is no d in Z such that d is a common divisor of y and z for all x in Q.

We prove that sqrt(2) isn't rational by contradiction. Let sqrt(2) be rational.
Let n in Z and let there be a dc\_251 in Z such that sqrt(2)*n=dc\_251 
and there is no dc\_255 in Z such that dc\_255 is a common divisor of n and dc\_251.
Let m in Z, let sqrt(2)*n=m and let there be no dc\_255 in Z such that dc\_255 is a common divisor of n and m.
N in Z, m in Z and sqrt(2)*n=m lead to 2*n\pow 2=m\pow 2.
Therefore m\pow 2 is even because n in Z and m in Z.
That implies that m is even because m in Z.
That implies that there is a dc\_263 in Z such that m=2*dc\_263.

Let k in Z and let m=2*k.
n\pow 2=2*k\pow 2 since n in Z, m in Z, k in Z, m=2*k and 2*n\pow 2=m\pow 2.
That implies that n\pow 2 is even since n in Z and k in Z.
That leads to even n because n in Z.
Hence 2 is a common divisor of n and m since m is even, n in Z and m in Z.
Thus we have a contradiction because there is no dc\_255 in Z such that no dc\_255 is a common divisor of n and m.

\end{minipage}}

\caption{Output of Omega Proverb}
\label{fig:naturallanguagesoutputomega}
\end{subfigure} 
\begin{subfigure}{0.69\textwidth}
\fbox{\begin{minipage}{0.99\linewidth}
\scriptsize
Let $x$ be an element of $f^{-1}(U)$. Then $f(x) \in U$.
Therefore, since $U$ is open, there exists $\eta > 0$ such that $u \in U$ whenever $d(f(x),u)< \eta$.
We would like to find $\delta > 0$ s.t. $y \in f^{-1} (U)$ whenever $d(x,y) < \delta$.
But $y \in f^{-1}(U)$ if and only if $f(y) \in U$.
We know that $f(y) \in U$ whenever $d(f(x),f(y)) < \eta$.
Since $f$ is continuous, there exists $\theta >0$ such that $d(f(x),f(y)) < \eta$ whenever $d(x,y) < \theta$.
Therefore, setting $\delta = \theta$, we are done.
\end{minipage}}
\caption{Output of Ganesalingam and Gowers problem solver}
\label{fig:naturallanguagesoutputGanesalingamGowers}
\end{subfigure} \\
\begin{subfigure}{0.99\textwidth}
\fbox{\includegraphics[width=0.99\linewidth]{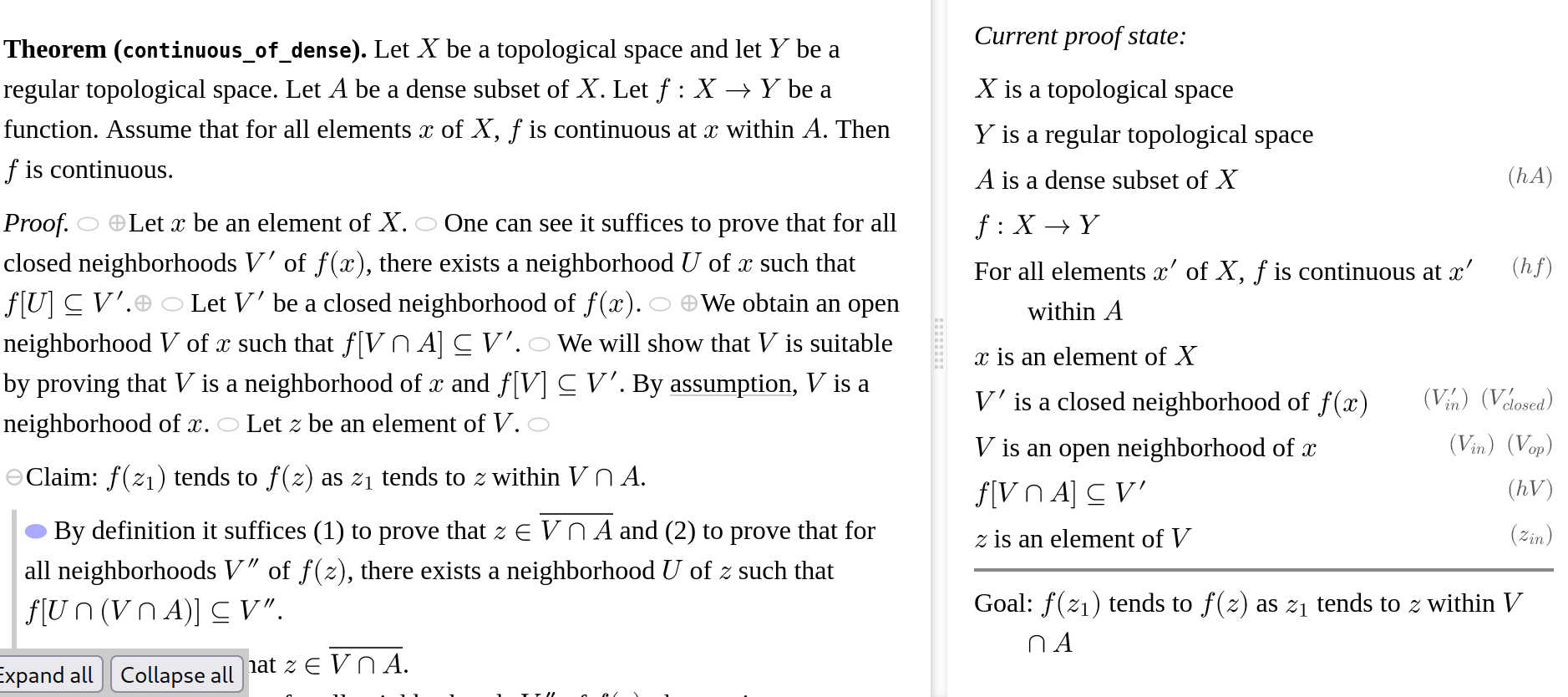}}
\caption{Output of Patrick Massot Lean-Html converter }
\label{fig:naturallanguagesoutputMassot}
\end{subfigure}

\caption{Natural Language Output}
\label{fig:naturallanguagesoutput}
\end{figure}

\subsection{Controlled Natural Languages (CNL)}

 The first step in this direction was the use of a declarative style~\cite{wiedijk-synthesis-2012} in place of procedural style as input in
 interactive theorem provers. While the latter consists of a sequence of imperative orders constructing the proof term 
 (the intermediate statements can generally be omitted or are considered as simple type descriptions; they can be computed and dynamically displayed by the software),
 the former expects the user to enter a list of statements, like in a traditional mathematical text, the proof terms justifying these statements
 being essentially automatically computed, even if the user can provide hints.
 
 Mizar~\cite{matuszewski-mizar-2005} is probably the first famous interactive theorem prover to accept a declarative style script.
 The Isabelle community followed when Wenzel developed the Isar extension in 2007~\cite{wenzel-isabelleisar-2007}. Figure \ref{fig:inputstyles} shows three simple proofs of the same statement, expressed in a proof-term language (in Lean), in a tactic procedural language (in Coq) and in a declarative style language (in Isabelle/Isar).

 In the 1960s, Glushkov carried out research aimed at designing a system featuring a low-level inference engine complemented by a high-level reasoner, 
 and parsing a powerful input language close to the natural mathematical language. These objectives stimulated the work of several research teams in Kiev in a program
 called ``Evidence Algorithm''~\cite{lyaletski-evidence-2010}. By 1990, a language (TL - Theory Language) and a parser and prover (SAD - System for Automated Deduction) were developed in a Russian version. This work was later taken up by Andriy Paskevych in Paris during his PhD thesis~\cite{rusinowitch-methodes-2007}. 
 TL evolved to ForTheL (Formula Theory Language), SAD was rewritten in English, the inference engine based on external ATPs, and theoretical advances enabled the development of the reasoner. Independently, Bonn, Schröder, Koepke, Kühlwein and Cramer were pursuing similar goals - a controlled natural language associated with an adapted prover~\cite{kuhlwein-naproche-2009,cramer-proof-checking-2013}. The Naproche system was developed during Cramer's PhD thesis in 2013 and later merged with SAD, enriching the language ForTheL and extending the parser to support LaTeX input~\cite{koepke-textbook-2019}. The program has been integrated into Isabelle IDE and benefits from
 the user interface and the access to the ATPs but do not share any logical connection with Isabelle prover~\cite{lon-isabellenaproche-2021}.

 On top of $\Omega$mega discussed above, Autexier and Wagner developed Plat$\Omega$, a ``mediator between text editors and proof assistance systems''~\cite{wagner-plat-2007}
which allows the user to enter her statements in a CNL named PL (proof language). Plat$\Omega$ converts the input into a format understandable by $\Omega$mega ;
in return, hints or completed proofs from $\Omega$mega are converted by Plat$\Omega$ into PL to be displayed by the editor (Texemacs).

  In ~\cite{xie-natural-2024}, the authors formally define a  ``natural-language-like proof language''
  modelling the structure of mathematics vernacular, with its syntactic ambiguities, and propose a tool to resolve them and check proofs written in the language.

  Lean Verbose is a controlled natural language in development since 2021 by Patrick Massot, implemented as a set of user-defined tactics in Lean4~\cite{kerjean-utilisation-2022,massot:LIPIcs.ITP.2024.27}. According to the author, using this tactic language is not easier to learn than the original language of Lean but using it improves students' performance while transitioning to handwritten proofs.

 With Waterproof~\cite{Wemmenhove-2024}  (developed since 2018), Jim Portegies and Jelle Wemmenhove imagined a custom tactic language on top of Coq that on the one hand requires the user to respect additional constraints (like type ascription, signposting, or distinguish ``take'' from ``assume''); but on the other hand,
 it relaxes the requirements on trivial steps justification. Note that a hint tactic is provided to suggest the introduction of the main connector of the goal and error messages have been refined. The software has been tested for four years at Eindhoven University of Technology with first-year undergraduate students.

 Figure \ref{fig:inputCNL} gives an overview of Diproche, LeanVerbose and Waterproof natural controlled languages.

\begin{figure}[!htb]
\begin{center}
  {\scriptsize
      \begin{minipage}[t]{1.11\linewidth}
      \begin{minipage}[t]{0.45\linewidth}
\begin{lstlisting}[style=lean]
/- Lean -/
lemma and_iff_of_imp :
  ∀ P Q : Prop, (P → Q) → ((P ∧ Q) ↔ P) :=
  
  λ P Q : Prop =>
    λ h_PimpQ : P → Q =>
      Iff.intro
      (
        λ h_PandQ : P ∧ Q =>
        (h_PandQ.left : P)
      )
      (
        λ h_P : P =>
          have h_Q : Q := h_PimpQ h_P
          (And.intro (h_P : P) (h_Q : Q) 
           : (P ∧ Q))
      )
\end{lstlisting}
       \end{minipage} \quad
       \begin{minipage}[t]{0.40\linewidth}
        \begin{lstlisting}[style=coq]
(* Coq/Roq *)
Lemma and_iff_of_imp :
  \forall P Q : Prop, (P -> Q) -> ((P /\ Q) <-> P).


  Proof.
    intros.
    constructor.
    intro H_PandQ.
    destruct H_PandQ as [HP HQ].
    exact HP.
    intro HP.
    constructor.
    - exact HP.
    - apply H.
      exact HP.
  Qed.

       \end{lstlisting}
       \end{minipage}
       \end{minipage}
       \begin{minipage}[t]{0.7\linewidth}
        \begin{lstlisting}[style=isabelle]
(* Isabelle/Isar *)
lemma and_iff_of_imp : "(P ⟶ Q) ⟶ ((P ∧ Q) ⟷ P)"
  proof
    assume h:"(P ⟶ Q)"
    have h1:"(P ∧ Q) ⟶ P"
    proof
      assume "P ∧ Q"
      thus "P"..
    qed
    (* cutting h2 *)
    from h1 and h2 show "((P ∧ Q) ⟷ P)" by blast
  qed
          \end{lstlisting}
       \end{minipage}
}
\end{center}
\caption{From top to bottom : Proof term language (Lean) ; Core tactic language (Coq) ; Declarative style language (Isabelle/Isar)}
\label{fig:inputstyles}
\end{figure}

\begin{figure}[!tb]

\begin{subfigure}{0.39\textwidth}
\fbox{\begin{minipage}[t]{0.9\linewidth}
\small Es sei x eine ganze Zahl. Zeige~: Wenn x gerade ist, dann is 2-3*x gerade.

Beweis:~: Es sei x gerade. Dann gibt es eine ganze Zahl mit x=2*k.
Sei k eine ganze Zahl mit x=2*k. \\
Dann is 2-3*x=2-3*(2*k)=2*(1-3*k).
Also ist 2-3*x gerade. 
qed.~\footnote{
Let $x$ be an integer. Prove: If $x$ is even, then $2-3x$ is even.

Proof: Let $x$ be even. Then, there is an integer $k$ such that $x=2k$.
Let $k$ be an integer with $x=2k$.
Then we have $2-3x=2-3\cdot(2k)=2(1-3k)$.
Hence $2-3x$ is even. qed.
}
\end{minipage}
}

\caption{Diproche language (german)}
\label{fig:inputCNLDiproche}
\end{subfigure}
\begin{subfigure}{0.59\textwidth}
\begin{lstlisting}[language=leanverbose,frame=single,basicstyle=\scriptsize\ttfamily]
Exercise "Continuity implies sequential continuity"
  Given: (f : ℝ → ℝ) (u : ℕ → ℝ) (x₀ : ℝ)
  Assume: (hu : u converges to x₀) (hf : f is continuous at x₀)
  Conclusion: (f ∘ u) converges to f x₀
Proof:
  Let's prove that ∀ ε > 0, ∃ N, ∀ n ≥ N, |f (u n) - f x₀| ≤ ε
  Fix ε > 0
  By hf applied to ε using that ε > 0 we get δ such that
    (δ_pos : δ > 0) and 
    (Hf : ∀ x, |x - x₀| ≤ δ ⇒ |f x - f x₀| ≤ ε)
  By hu applied to δ using that δ > 0 
    we get N such that Hu : ∀ n ≥ N, |u n - x₀| ≤ δ
  Let's prove that N works : ∀ n ≥ N, |f (u n) - f x₀| ≤ ε
  Fix n ≥ N
  By Hf applied to u n it suffices to prove |u n - x₀| ≤ δ
  We conclude by Hu applied to n using that n ≥ N
QED
\end{lstlisting}
\caption{LeanVerbose: Lean dialect}
\label{fig:inputCNLLeanVerbose}
\end{subfigure}
\\
\begin{center}
\begin{subfigure}{0.59\textwidth}
\begin{lstlisting}[style=coq,basicstyle=\scriptsize\ttfamily]
Lemma exercise_1 : 2 is the infimum of [2,5).
Proof.
We need to show that
  (2 is a lower bound for [2,5) ∧
    (for all m : ℝ, m is a lower bound for [2, 5) ⇒  m  ≤  2)).
We show both statements.
- We need to show that (2 is a lower bound for [2, 5)).
  We need to show that (for all x : ℝ, x : [2, 5) ⇒  2  ≤  x).
  Take x : ℝ. Assume that (x : [2, 5)).
  We conclude that (2 ≤ x).
- We need to show that
    (for all m : ℝ, m is a lower bound for [2, 5) ⇒  m  ≤  2).
  Take m : ℝ. Assume that (m is a lower bound for [2, 5)).
  It holds that (2 : [2, 5)).
  We conclude that (m ≤  2).
\end{lstlisting}
\caption{Waterproof: Coq dialect}
\label{fig:inputCNLWaterproof}
\end{subfigure}
\end{center}
\caption{Input in controlled natural languages}
\label{fig:inputCNL}
\end{figure}

\section*{Conclusion}

We have presented an overview of the use and design of proof assistants for teaching.
Furthermore, we highlighted the fact that proof assistants used for teaching are diverse and reflect different teaching objectives and contexts.
 They differ in the design (based on a general purpose proof assistant or specialized for teaching), the proof language and mode of interaction,
 the visualization of proofs and feedback provided to the students.
Proof assistants have been used for a long time for teaching, and many tools have been designed. In some fields such as foundations of software and logic, some teaching material has been produced and shared, and the use of proof assistants is widespread. 
However, the evaluation of the impact of these tools from a didactic point of view and their use for teaching mathematics is still at its debut.
Proof assistants can be used for various purposes, targeting very different students under the supervision of teachers with different skills and goals. They can be used for teaching the concept of proof  as well as logic or advanced mathematics or computer science.  This is why the evaluation of these tools is a complex issue, and it is difficult to define a unique set of goals for such an evaluation. Recently, Keenan and Omar have proposed design criteria and evaluation methods~\cite{keenan-learner-centered-2024}. %
A growing community of researchers is addressing these issues, and we are optimistic that using proof assistants for both teaching and research will be commonplace by the middle of the 21$^{th}$ century.

\paragraph{Acknowledgments}

This work is partially funded by the ANR Project APPAM ANR-23-CE38-0009-01.

\clearpage

\bibliographystyle{eptcsalpha}
\bibliography{biblio-clean} 

\end{document}